\documentclass[aps,prl,twocolumn,superscriptaddress,floatfix,citeautoscript]{revtex4}
\usepackage{times}
\usepackage{graphicx}
\usepackage{dcolumn}
\usepackage{bm}
\usepackage{color}
\usepackage{amsmath}
\usepackage{amssymb}
\newcommand {\beq} {\begin{equation}}
\newcommand {\eeq} {\end{equation}}
\newcommand {\bqa} {\begin{eqnarray}}
\newcommand {\eqa} {\end{eqnarray}}
\usepackage[colorlinks=true]{hyperref}

\begin{document}

\title{\textbf{Enigma of the vortex state in a strongly correlated d-wave superconductor}}

\author{Anushree Datta\footnote{Present Address: Universit\'e Paris Cit\'e, Laboratoire Mat\'eriaux et Ph\'enomenes Quantiques, CNRS, F-75013, Paris, France, Universit\'e Paris-Saclay, CNRS, Laboratoire de Physique des Solides, F-91405, Orsay, France.}}
\affiliation{Indian Institute of Science Education and Research Kolkata, Mohanpur, India-741246}

\author{Hitesh J. Changlani}
\affiliation{Department of Physics, Florida State University, Tallahassee,  Florida 32306, USA}
\affiliation{National High Magnetic Field Laboratory, Tallahassee, Florida 32310, USA}

\author{Kun Yang}
\affiliation{Department of Physics, Florida State University, Tallahassee, Florida 32306, USA}
\affiliation{National High Magnetic Field Laboratory, Tallahassee, Florida 32310, USA}

\author{Amit Ghosal}
\affiliation{Indian Institute of Science Education and Research Kolkata, Mohanpur, India-741246}
\begin{abstract}
We show that strong electronic repulsion transforms a vortex core from a metallic-type in overdoped regime to a Mott-insulator at underdoping of a strongly correlated d-wave superconductor. This changeover is accompanied by an accumulation of electron density at the vortex core towards local half-filling in the underdoped region, which in turn facilitates the formation of the Mott insulating core. We find that the size of vortices evolves non-monotonically with doping. A similar non-monotonicity of critical field $H_{\rm c2}$, as extracted from superfluid stiffness, is also found. Our results explain some recent experimental puzzles of cuprate superconductors.
\end{abstract}
\maketitle

\medskip

\noindent {\it \bf Introduction.} Topological defects, such as, vortices have drawn significant research interests ever since Kosterlitz and Thouless~\cite{Kosterlitz_1973,RevModPhys.89.040501} established a melting mechanism mediated by them. Vortices are low-lying excitations of type-II superconductors in the presence of magnetic fields. In conventional superconductors, a magnetic field produces a periodic array of vortices~\cite{AVL,Book2} with a normal metallic core of size $\xi$ with circulating currents around the vortex on the scale of the penetration depth $\lambda$~\cite{Tinkham}. With increasing field $H$, the density of vortices increases. Beyond the critical field $H_{\rm c2}$, overlapping cores suppress pairing amplitude everywhere and the superconductor transitions into a metal~\cite{AVL}. The study of vortices in unconventional superconductors has gathered recent momentum due to several experimental puzzles~\cite{Yin2015,PhysRevX.11.031040,doi:10.1126/science.aat1773}.

One such mystery lies in the mapping of local density of states (LDOS) at the vortex core in cuprate superconductors, a prototype of strongly correlated d-wave superconductors (dSC). Differential conductance in cuprates (both in ${\rm YBa_{2}Cu_{3}O_{7-\delta}}$~\cite{PhysRevLett.75.2754} and ${\rm Bi_{2}Sr_{2}CaCu_{2}O_{8+\delta}}$~\cite{PhysRevLett.85.1536}) in optimal to underdoping region shows a gap structure, while weak-coupling calculations predict a large accumulation of low-lying states in LDOS at vortex core for all dopings, $\delta$~\cite{PhysRevB.52.R3876}. Recent experiments find similar significant pileup of the low-lying states but in the overdoped regime~\cite{PhysRevX.11.031040}. Several theoretical attempts have been made to understand the low doping anomalous behaviours~\cite{PhysRevB.74.144516,PhysRevB.76.094509,PhysRevB.68.012509,PhysRevB.63.134509}, including the generation of sub-dominant competing orders at vortex core, such as antiferromagnetic~\cite{PhysRevLett.87.147002,PhysRevB.66.214502,PhysRevB.66.094501}, s-wave pairing~\cite{doi:10.1143/JPSJ.66.3367}, d-density wave~\cite{PhysRevB.68.024513,PhysRevB.76.020511} and pair-density wave orders~\cite{PhysRevB.97.174511}, augmented to weak-coupling descriptions. However, no consensus has yet been achieved to comprehend the anomaly~\cite{Bruer2016,PhysRevX.11.031040}. The role of strong correlations on the vortex inhomogeneities, however, have largely alluded the field of research, see however,~\cite{PhysRevB.66.094513,PhysRevLett.87.167004,PhysRevLett.85.1100}. After all, these strong electronic repulsions turn the parent undoped ($\delta=0$) compound an antiferromagnetic Mott insulator~\cite{RevModPhys.78.17}.

Taking the route of direct inclusion of strong correlations by removing any double-occupancy within a fully self-consistent microscopic calculation, our main results in this paper are: (i) Underdoped d-wave vortex state induces charge accumulation towards local half-filling at the vortex core, and thereby promotes the emergence of `Mottness'. (ii) The changeover of the nature of the vortex core from being Mott insulating to metallic with increasing doping, which explains the tunneling spectroscopic measurements of LDOS. (iii) The size of vortices show intriguing non-monotonic behavior. Such a non-monotonic behavior has other fascinating implications. For example, our result of superfluid density in the presence of magnetic field indicates that the upper critical field $H_{\rm c2}$, shows a dome shaped evolution with $\delta$, in agreement with experimental findings.

\noindent {\it \bf Model and methods.} 
Strongly correlated materials can be described minimally by the Hubbard model~\cite{SCALAPINO1995329} with $U \gg t$. In this limit the low energy physics is described by a $t-J$ model~\cite{Anderson_2004}:

\begin{eqnarray}
{\cal H_{\rm{t-J}}}&=& -t\sum_{\langle ij\rangle \sigma} {\cal P}\left(e^{i\phi_{ij}}{\hat{c}}^{\dagger}_{i\sigma}{\hat{c}}_{j\sigma} + \rm{H.c.}\right) {\cal P}- \sum_{i}\mu \hat{n}_{i}\nonumber\\ &+& J\sum_{\langle ij\rangle} {\cal P}\left(\mathbf{{{S}}_{i}}\cdot\mathbf{{{S}}_{j}}- \frac{\hat{n}_{i}\hat{n}_{j}}{4}\right){\cal P}.
\label{tj}
\end{eqnarray}
Here $\hat{c}^{\dagger}_{i\sigma}$ ($\hat{c}_{i\sigma}$) is the creation (annihilation) operator of an electron with spin $\sigma$ at lattice site $i$ in a two-dimensional square lattice, $\mathbf{S}_{i}$ and $\hat{n}_i$ are the spin and electron density operators, respectively, $\langle ij\rangle$ denotes nearest neighbor bonds, $t$ is the hopping amplitude for an electron to its nearest neighbors, $\mu$ is the chemical potential fixing the average electron density $\rho$, $J = 4t^{2}/U$ is the super-exchange interaction with $U$ being the onsite Hubbard repulsion strength. Here, ${\cal P}$ is the projection operator which prohibits double occupancies on each lattice site due to the strong onsite repulsive $U$. The orbital magnetic field is incorporated through the Peierls factor: $\phi_{ij} = {\pi/\phi_{0}} \int^{j}_{i} \mathbf{A}.\mathbf{dl} $, where $\phi_{0} = hc/2e$ is the superconducting (SC) flux quantum. We consider a uniform orbital field $\mathbf{H} = H \hat{z}$ and choose to work with the Landau gauge, $\mathbf{A} = Hx\hat{y}$.

The effect of the projection operator is implemented by the Gutzwiller approximation (GA)~\cite{PhysRevB.76.245113}, where restriction of double occupancy is removed in expense of renormalizing the hopping and exchange parameters: $t_{ij} \rightarrow g^{t}_{ij} t$, $J_{ij} \rightarrow g^{J}_{ij} J$, here $g$'s are the corresponding Gutzwiller renormalization factors (GRFs)~\cite{PhysRevB.76.245113,FCZhang_1988}. GRFs, which depend on local densities $n_{i}$, are provided in the Supplementary Material (SM)~\cite{fnSM}. Physically, the removal of double occupancy prohibits certain hopping processes across the bond $\langle ij \rangle$, and hence the average kinetic energy must reduce on that bond from a situation where double occupancies are allowed. This is incorporated by the hopping GRFs $g^{t}_{ij} \le 1$. Similarly, the overall higher probability of sites being singly occupied enhances the exchange coupling through $g^{J}_{ij}$.
The GA formalism has been verified~\cite{PhysRevLett.98.027004, PhysRevB.60.R9935} to agree well with variational Monte Carlo calculations~\cite{PhysRevLett.87.217002} (where the projections are exact) for homogeneous systems. We note that we refer to the strong correlations equivalently with the removal of double occupancy in this work.  

We take advantage of the perfect periodicity of our square vortex lattice~\footnote{While a triangular vortex lattice is energetically favorable 
within a continuum Ginzburg-Landau theory, which ignores underlying lattice symmetries. However, it is numerically challenging to study the triangular vortex lattice with a underlying square lattice of finite size. Also, the connection between the structure of the vortex lattice and the crystal lattice symmetry is observed experimentally in conventional s-wave superconductors~\cite{Ganguli_2016}} by solving the eigenvalue problem using a fully self-consistent Bogoliubov de-Gennes (BdG) method on a unit cell typically of size $N=24\times 48$ and then extending the wavefunction on a system made of typically $16 \times 8$ unit cells~\cite{PhysRevB.52.R3876,PhysRevB.66.214502}. We present all energies in units of the hopping amplitude $t$ and set the temperature $T=0$ for our calculations. We use $J = 0.33$ – a typical value used for cuprate superconductors~\cite{PhysRevB.52.615}. We consider several doping ($\delta=1-\rho$) values ranging from $\delta= 0.06$ (underdoped) to $\delta= 0.25$ (overdoped). To emphasize our key findings, we compared our results from Gutzwiller inhomogeneous mean-field theory (GIMT) with results from standard inhomogeneous mean-field theory (IMT), where the effects of projection $\cal P$ are ignored by taking the Gutzwiller factors to be unity, i.e., with double occupancy being allowed. In IMT, we tune $J$ values for each doping in such a way that both IMT and GIMT yield the same d-wave gap when the magnetic field is zero~\cite{Garg2008}. The details of GIMT and IMT calculations are included in the SM~\cite{fnSM}.

\begin{figure}
\centering
    \includegraphics[width=0.49\textwidth]{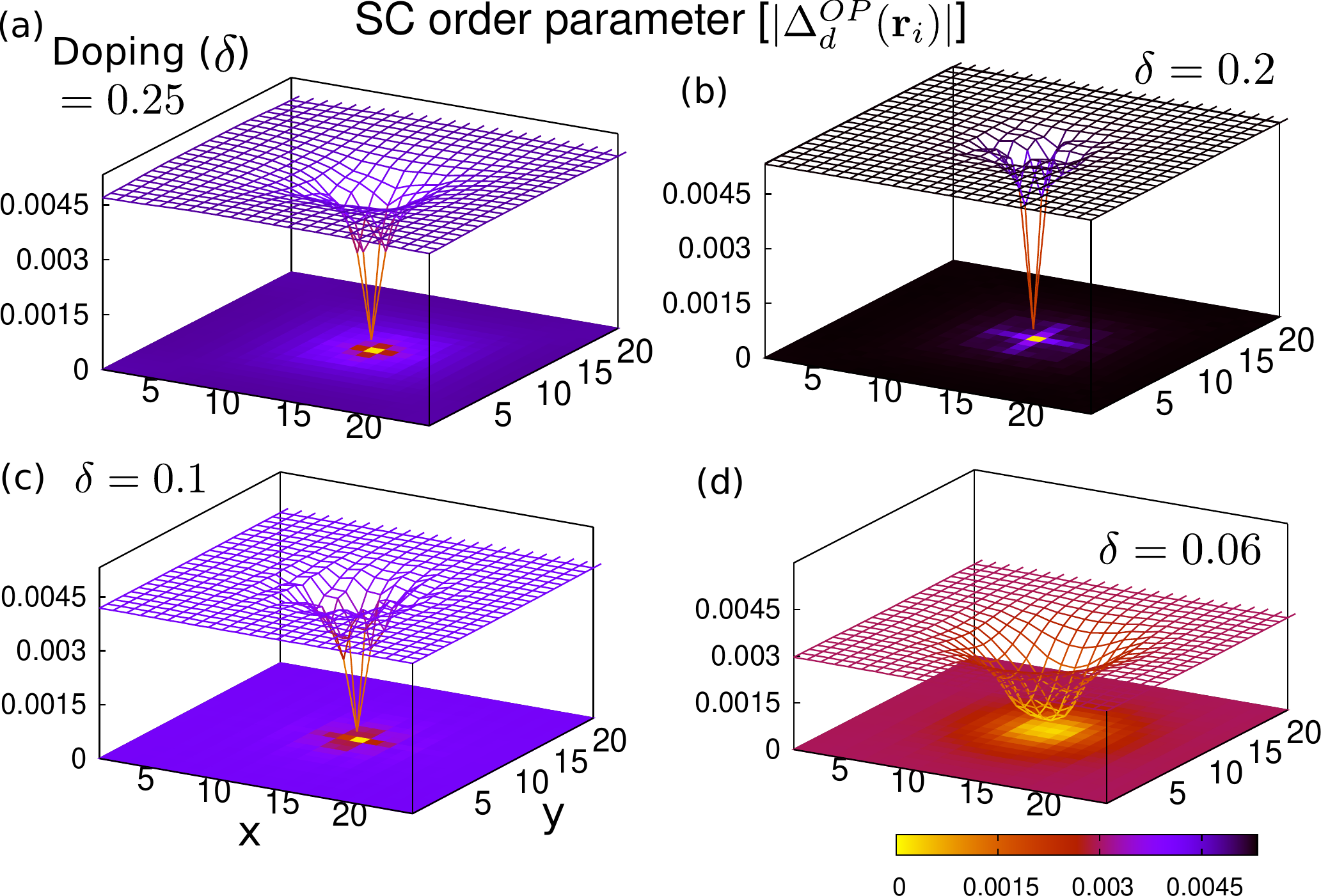}
\caption{{\bf SC order parameter profiles.} d-wave SC $|\Delta^{\rm OP}_{\rm d}(\bf{r}_{i})|$ profiles around a vortex core on a magnetic cell of size $24\times24$ at different doping ($\delta$) values. The fall of $\Delta^{\rm OP}_{\rm d}(\bf{r}_{i})$ at the vortex center has the conventional conical form at $\delta=0.25$, $0.2$, and takes up a form of a ``flat-bottom bowl" at $\delta=0.06$. }
\label{fig:fig1}
\end{figure}

\noindent {\it \bf d-wave SC order.} We begin describing our results by elaborating on the dSC order parameter calculated within the GIMT framework: $\langle\hat{c}_{i\sigma}\hat{c}_{j\overline{\sigma}}\rangle_{\psi}\approx g^{t}_{ij}\Delta_{ij}$\cite{PhysRevB.78.115105,PhysRevB.95.014516}. Here $\langle ..\rangle_{\psi}$ denotes the expectation value in the truncated Hilbert space with double occupancies removed. The spatial profile of the dSC order parameter, $\Delta^{\rm OP}_{\rm d}({\bf r}_{i})=\frac{J}{4} |[g^{t}_{i,i+\hat{x}}\Delta_{i, i+ \hat{x}}+ g^{t}_{i,i-\hat{x}}\Delta_{i, i-\hat{x}}-e^{ibx}g^{t}_{i,i+\hat{y}}\Delta_{i, i+\hat{y}}-e^{-ibx}g^{t}_{i,i-\hat{y}}\Delta_{i, i-\hat{y}} ]|$ (here $b\equiv H/\phi_{0}$) is shown in Fig.~\ref{fig:fig1}. Different panels of Fig.~\ref{fig:fig1} show $\Delta^{\rm OP}_{\rm d}({\bf r}_{i})$ for representative $\delta$. Away from a vortex core, i.e. near the boundary of the magnetic unit cell containing a single SC flux quantum, $\Delta^{\rm OP}_{\rm d}$ attains the homogeneous Bardeen-Cooper-Schrieffer (BCS) value while it falls at the vortex core. This conical-shaped fall at the core for overdoped [Fig.~\ref{fig:fig1}(a)] to optimally doped [Fig.~\ref{fig:fig1}(b, c)] systems follows the expected ${\rm tanh}(r/\xi)$ behavior, where $\xi$ is the SC coherence length~\cite{Tinkham}. In contrast, the fall of $\Delta^{\rm OP}_{\rm d}({\bf r}_{i})$ shows a strikingly different pattern at underdoping [Fig.~\ref{fig:fig1}(d)]: The region of the depletion of $\Delta^{\rm OP}_{\rm d}({\bf r}_{i})$ is much wider -- near the core-center, the vortex resembles a ``flat-bottom bowl". The weak-coupling IMT calculations preserve the conical-shaped vortex for all $\delta$, and shrinks monotonically towards underdoping, see SM~\cite{fnSM}. 

\begin{figure}
\centering
    \includegraphics[width=0.49\textwidth]{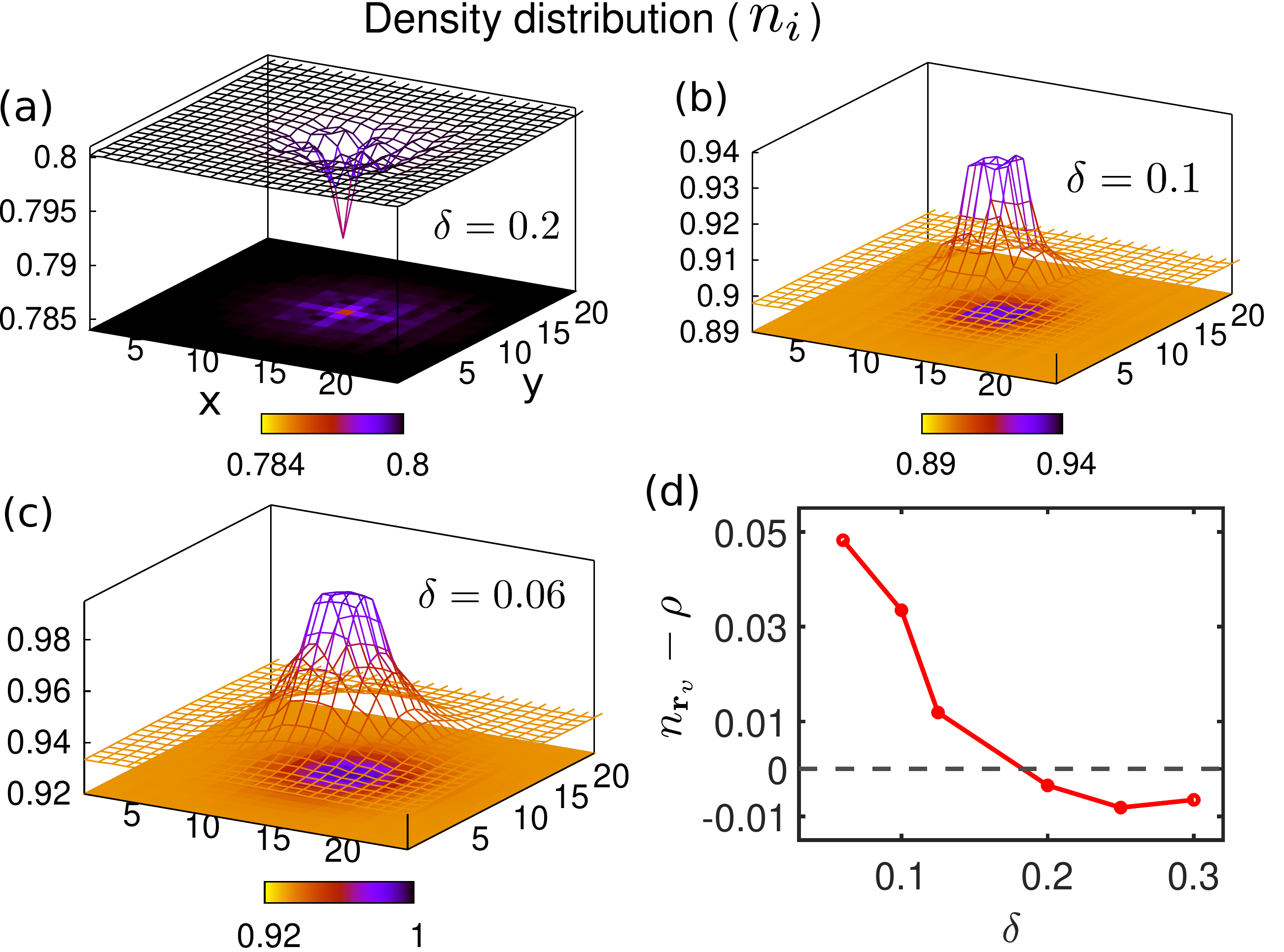} 
\caption{{\bf Electronic charge density distribution.} Local density $n_{i}$ maps around a vortex core for different $\delta$. At the vortex core, at $\delta=0.2$ [panel (a)], $n_{i}$ features a dip, and at $\delta=0.1$, $0.06$ (panel (b) and (c)]), the electronic charges accumulate to form a hill (with core density approaching unity). Panel (d) shows the profile of $n_{\rm {\bf r}_v} - \rho$ Vs $\delta$. The $n_{{\bf r}_v}$ values are less than $\rho$ for $\delta>0.18$ and greater than $\rho$ for $\delta< 0.18$. At $\delta=0.06$, $n_{\rm {\bf r}_v}$ approaches unity leading to formation of a Mott insulating core. 
}
\label{fig:fig2}
\end{figure}

\noindent {\it \bf Local charge density at a vortex core.} In order to develop a deeper insight into above results we next study the local charge density near the vortex core location, ${\bf r}_v$~\footnote{While ${\bf r}_v$ represents the center of a vortex, for a better resolution of different local observables, e.g. LDOS at vortex core, we gather statistics not just at the vortex center but on a $2 \times 2$ lattice sites around the vortex center. Thus ${\bf r}_v$ represents the location of the `vortex core region'.} for different $\delta$. In the optimally doped region ($\delta=0.2$), the spatial density profile features a weak dip around ${\bf r}_v$ [Fig.~\ref{fig:fig2}(a)], consistent with the weak-coupling theory. Upon lowering $\delta$, the $n_{{\rm {\bf r}_v}}$ rises rapidly to near unity by $\delta=0.06$ [Fig.~\ref{fig:fig2}(c)]. This enhancement of $n_{{\rm {\bf r}_v}}$ characterizes the emergence of `Mottness' at the vortex core region for an average doping not so close to unity. Thus, for $\delta \lesssim 0.06$ the vortex core becomes insulating and $g^{t}_{ij}\approx 0$ quenching the kinetic energy at the core. The effective picture of the underlying normal state in the core becomes that of an undoped patch of (antiferromagnetic) Mott insulator, described by a local Heisenberg model. This is quite unlike the Abrikosov vortex with a metallic core~\cite{Tinkham}. We note that the vortex core here is not simply serving as a window to the underlying normal state in the sense that the underlying normal state at $\delta=0.06$ without the vortex is not yet a Mott-insulator. Instead, the Mott vortex core here is a result of strong correlations and a by-product of charge accumulation due to it. However we should also emphasize that this limit of vortex core is realized only in the proximity of the undoped Mott insulator. The reorganization of the local charge density at the vortex core as a function of doping is shown in Fig.~\ref{fig:fig2}(d). We find the excess local charge density at the vortex core changes sign with $\delta$ near optimal doping.

The non-linear effects of GRFs in the effective chemical potential $\mu_{i}$, obtained while minimizing the total ground state energy of the system, play a key role in driving vortex cores towards Mottness, see SM for additional details~\cite{fnSM}. Such effects not only drive the vortex core towards Mottness but also helps the nearby sites of the vortex core to attain local half-filling forming a near plateau in $n_i$ [Fig.~\ref{fig:fig2}(c)]. The occurrence of a plateau in $n_{i}$ in the core region is ultimately connected to the ``flat bottom bowl" structure of $\Delta^{\rm OP}_{\rm d}$. The charge fluctuations freeze on these sites, as $t_{ij}\approx 0$, depleting dSC order over an extended region. 

We emphasize that the removal of double occupancy is crucial for the aforementioned charge accumulation at the core and subsequent effects. Without the removal of double occupancy, we verified that the weak dip in $n_{i}$ at the vortex core, a feature of overdoping continues until the lowest doping, see SM~\cite{fnSM}.

\noindent {\it \bf LDOS at the vortex core.} The emergence of Mottness has important implications for the LDOS at the vortex core as we discuss below. In an s-wave superconductor, Andreev-like zero-energy bound states~\cite{CDGM_PhysLett9_307} were predicted theoretically to appear in the vortex cores and have also been observed experimentally in tunneling measurements~\cite{PhysRevLett.62.214}. For a dSC, similar accumulation of the low-energy core states (LECS) is also predicted within IMT calculation~\cite{PhysRevB.52.R3876}, even though true bound states are not found due to the collapse of the d-wave gap along the nodal directions. Such LECS are reminiscent of the metallic nature of the vortex core. However, the differential tunneling conductance map in cuprates shows no signatures of LECS in underdoped to optimally doped samples, beyond some sub-gap features~\cite{PhysRevLett.75.2754,PhysRevLett.85.1536}. In contrast, recent experiments in overdoped samples showed prominent LECS at the vortex core~\cite{PhysRevX.11.031040}.

\begin{figure}
\centering
    \includegraphics[width=0.495\textwidth]{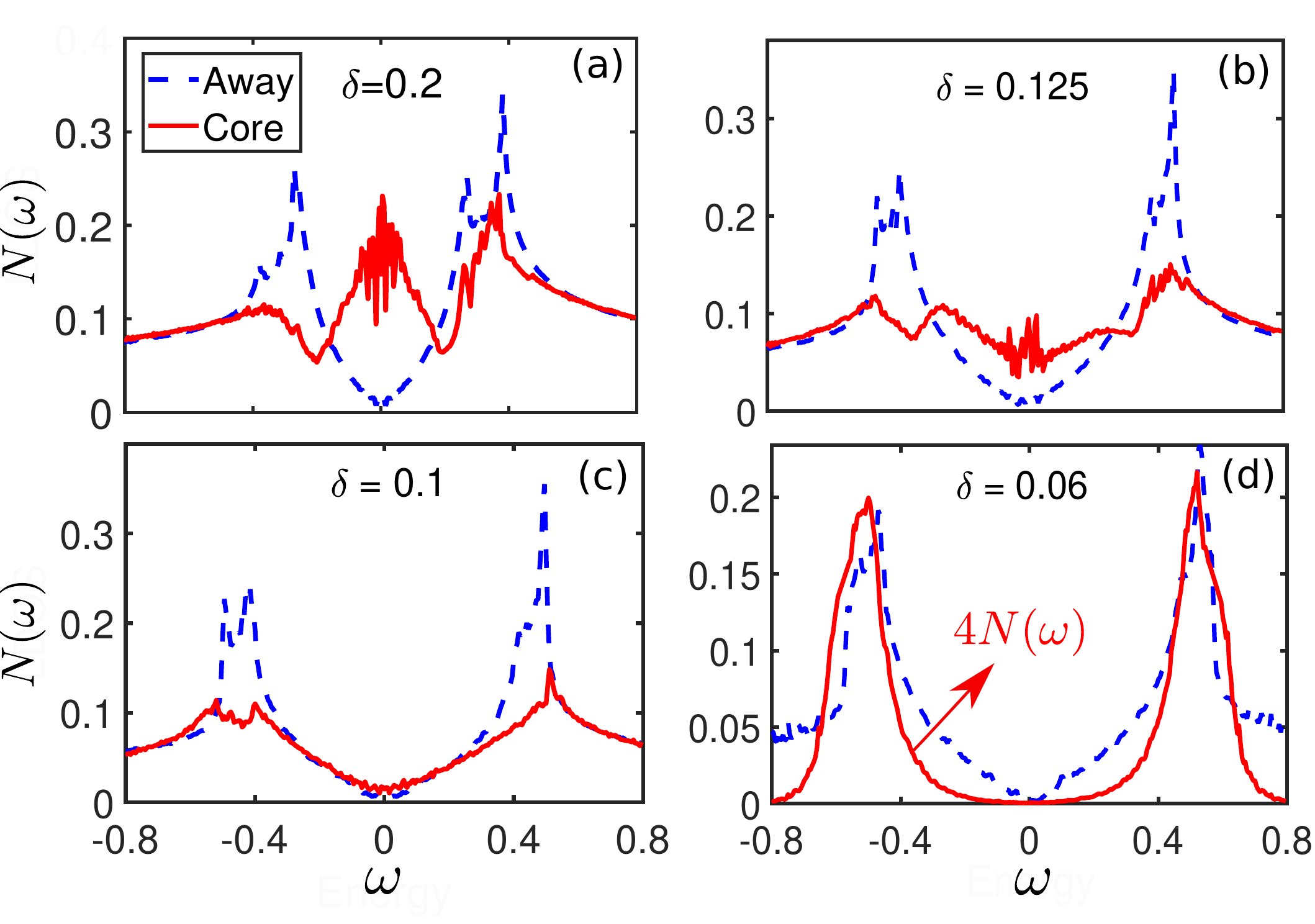} 
  \caption{{\bf Local density of states.} LDOS at the vortex core (red traces) and away from the core (blue traces) for $\delta=0.2$ (a), $\delta=0.125$ (b), $\delta=0.1$ (c), and $\delta=0.06$ (d). For $\delta=0.2$, the LDOS features a mid-gap peak which gradually reduces with decreasing $\delta$. For $\delta=0.06$, a hard gap opens with sharp peaks at $\omega\approx \pm J_{\rm eff}/2$. In panel (d), the vortex core LDOS is scaled up by a factor of $4$, for visual clarity.}
 \label{fig:fig3}
\end{figure}

To uncover this mystery, we show in Fig.~\ref{fig:fig3} the LDOS with varying doping $\delta$ in GIMT. Within GIMT, the LDOS is calculated using~\cite{Garg2008,PhysRevB.100.035114}: $N({\bf r}_{i},\omega)=N^{-1}_{e} \sum_{k,n} g^{t}_{ii}[ |u^{k}_{n}({\bf r}_{i})|^{2} \delta\left(\omega-E_{k,n}\right)+ |v^{k}_{n}({\bf r}_{i})|^{2} \delta\left(\omega+E_{k,n}\right)]$, where~$\{u^{k}_{n}({\bf r}_{i}), v^{k}_{n}({\bf r}_{i})\}$~are the local Bogoliubov wavefunctions, $E_{k,n}$ are corresponding energy eigenvalues (see SM~\cite{fnSM}), and $N_e$ is the total number of eigenstates. As shown Fig.~\ref{fig:fig3}(a) the LDOS near the vortex cores is found to feature a peak near zero-energy for optimal doping $\delta=0.2$. We find a similar peak at $\omega\approx 0$ in LDOS near vortex core for doping $\delta>0.2$. Thus, LECS are present in the overdoped to optimally doped region, which also agrees with the weak-coupling predictions~\cite{PhysRevB.52.R3876}. However, the vortex core LDOS at $\delta=0.125$ in Fig.~\ref{fig:fig3}(b) shows a depletion in zero energy states and subgap features. With decreasing doping the low energy states get further suppressed and no LECS can be seen in Fig.~\ref{fig:fig3}(c). Upon further lowering doping to $\delta=0.06$, the vortex core LDOS exhibits a U-shaped (hard) gap, as depicted in Fig.~\ref{fig:fig3}(d). This gap can be explained by the change in the nature of the vortex core with core density approaching unity for $\delta=0.06$ as seen in Fig.~\ref{fig:fig2}(c). The Mott cluster of sites at the vortex core, being described by an effective Heisenberg model as discussed already, features lowest lying excited states beyond a spin gap $\approx J_{\rm eff}$~\cite{PhysRevB.95.014516, PhysRevLett.96.017001}. The tantalizing similarity of our finding of LDOS with experiments is truly intriguing. In IMT calculations, prominent LECS are always present at the vortex core for all $\delta$, see SM~\cite{fnSM}.

\noindent {\it \bf Non-monotonicity in the core size.} The unfolding of Mottness causes an intriguing non-monotonic variation of the core size with $\delta$, as we examine below.

\begin{figure}
\centering
    \includegraphics[width=0.45\textwidth]{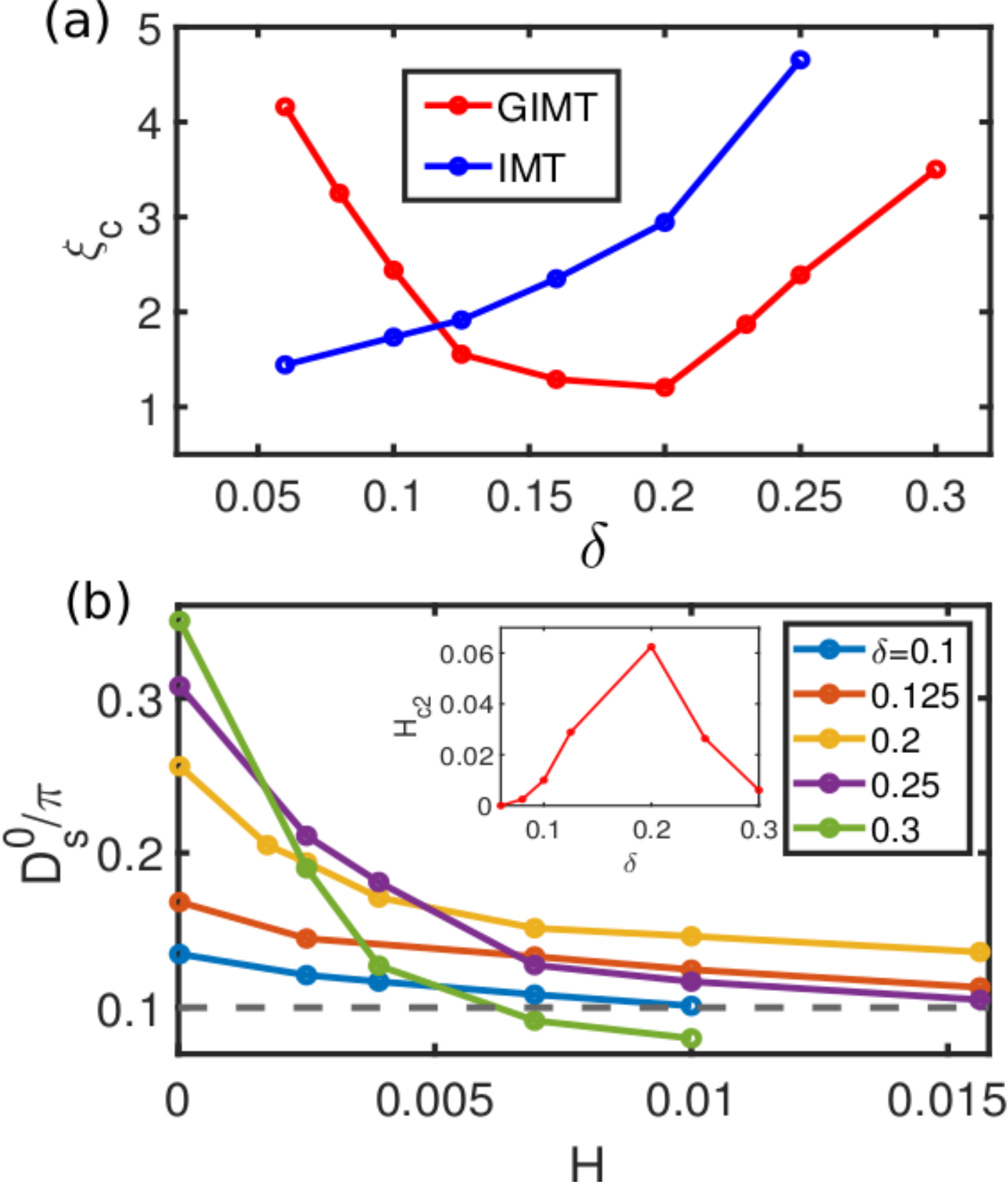} 
  \caption{{\bf Vortex core size and critical magnetic field from superfluid density.} Panel (a) depicts the variation of the vortex core length scale $\xi_{\rm c}$ as a function of doping, from IMT (blue trace) and GIMT (red trace) calculations. In IMT, $\xi_{\rm c}$ shrinks monotonically with decreasing doping. In GIMT, $\xi_{\rm c}$ shows a non-monotonic behavior. Values of $\xi_{\rm c}$ are in the unit of the lattice spacing. Panel (b) shows the variations of superfluid density $D^{0}_s$ as a function of magnetic field $H$ at different doping values. The threshold value for estimating the critical magnetic field $H_{\rm c2}$  is set at $D^{0}_s/\pi = 0.1$, as marked by
the black horizontal line. The inset in panel (b) shows the behavior of the obtained $H_{\rm c2}$ with respect to $\delta$, featuring a dome like profile. The $H$ values are represented in the unit of $\phi_{0}$.
}
\label{fig:fig4}
\end{figure}

For definiteness, we define the vortex length scale $\xi_{\rm c}$ as the distance from the vortex center where the order parameter $\Delta^{\rm OP}_{\rm d}(i)$ recovers $80\%$ of its maximum value. 
The red trace in Fig.~\ref{fig:fig4}(a), representing $\xi_{\rm c}(\delta)$, captures the two trends above and below the optimal doping $\delta \approx 0.2$.  For $\delta>0.2$, $\xi_{\rm c}$ shrinks as the doping value is decreased. This is consistent with the BCS expectation, where $\xi_{\rm c}\sim v_{\rm f}/\pi E_{\rm gap}$, with $v_{\rm f}$ and $E_{\rm gap}$ being the Fermi velocity and the energy-gap, respectively. Since, $E_{\rm gap}$ increases with decreasing $\delta$ within a d-wave BCS description, the vortex core shrinks. In the region below $\delta\approx0.2$, $\xi_{\rm c}$ ceases to follow the $v_{\rm f}/\pi E_{\rm gap}$ trend and starts to increase continuously as doping is lowered towards $\delta \rightarrow 0$. As discussed earlier, in the strong underdoped limit the congregation of Mott sites makes the variation of $\Delta^{\rm OP}_{\rm d}$ near the vortex core flatter. Our findings indicate that the enhancement of $\xi_{\rm c}$ in underdoped regime is intimately connected with formation of Mott-cluster.
It is indeed fascinating that the non-monotonicity in the vortex state tracks the non-BCS behavior~\cite{RevModPhys.78.17}. A similar non-monotonic doping dependence has been theoretically discussed also for the SC coherence length in strongly correlated superconductors~\cite{PhysRevLett.87.217002}.

To further highlight the prominent dependence of the vortex core size on strong correlations, we also include the trace of $\xi_{\rm c}$ from IMT calculations in Fig.~\ref{fig:fig4}(a), which shows only a monotonic increase with $\delta$ in the entire range. 

\noindent {\it \bf Superfluid stifness and critical magnetic field.} Having encountered the non-monotonic dependence of $\xi_{\rm c}$ with $\delta$, we next turn our attention to superfluid stiffness $D_{s}$ which gives rise to Meissner effect~\cite{Tinkham}. Here we focus on the $\delta$-dependence of $H_{\rm c2}$ within GIMT framework. In what follows, we calculate $D_{s}$ using the Kubo formalism~\cite{PhysRevB.47.7995}: $D_{s}/\pi=\langle -k_{x}\rangle -\Lambda_{xx}\left(q_{x}=0, q_{y}\rightarrow 0, \omega_{n}=0\right)$,~where $\langle -k_{x}\rangle$ is the average kinetic energy along $x$-direction and $\Lambda_{xx}(\mathbf{q},\omega)$ is the transverse current-current correlation function. In order to obtain the $H_{\rm c2}(\delta)$, in Fig.~\ref{fig:fig4}(b) we plot $D_{s}$ as a function of $H$, at different values of $\delta$. Because the BdG technique does not include quantum phase fluctuations of SC order, $D_{s}$ is not driven to zero by the fluctuations in the dSC pairing amplitude alone (which are fully included in BdG method). However, because BdG calculation results in a significant reduction of $D_{s}$ to a low value, it is expected that quantum phase fluctuations, riding on top of the fluctuations in the pairing amplitude, would guide $D_{s}$ to zero. We thus consider a small threshold value of $D_{s}/\pi=0.1$ to mark off $H_{\rm c2}$. Even though such extraction of $H_{\rm c2}$ will not be an accurate estimate of the upper critical fields, we believe it to represent the qualitative doping dependence of the {\it true} $H_{\rm c2}$.

The behavior of the extracted critical field $H_{\rm c2}$ in the inset of Fig.~\ref{fig:fig4}(b), features a dome-shaped profile with its maximum residing at $\delta\approx 0.2$ (optimal doping). Similar non-monotonic behavior in $H_{\rm c2}$ versus $\delta$ has been recently observed in cuprate superconductors~\cite{Wen_2003}. Interestingly, this finding gels well with the size of vortex core, because in Ginzburg – Landau theory theories, $H_{\rm c2}=\phi_{0}/2\pi\xi^{2}$, where the coherence length $\xi$ is the characteristic length scale of the vortex core. Thus a non-monotonicity in the core size, as seen in Fig.~\ref{fig:fig4}(a), implies a non-monotonicity in ${H}_{\rm c2}$ as well.  Interestingly, in cuprates the maximum of $H_{\rm c2}$ occurs near the optimal doping~\cite{PhysRevB.86.174501,Wang2008}, similar to our findings.

\noindent{\it \bf Conclusion.} 
We illustrated how the nature of the vortex core changes from metallic-type in overdosed regime to a Mott-insulating one upon approaching undoping of a strongly correlated dSC. This changeover is accompanied by accumulation of the electronic charge at vortex core towards half-filling, which in turn facilitate the formation of Mott insulating core. It will be interesting to track the charge of vortices using cavity electromechanics measurements~\cite{doi:10.1021/acs.nanolett.1c04688}. The change of the nature of vortex explains the anomaly in LDOS with dopings. The shape of the vortices do change as well, leading to a non-monotonic evolution of the vortex core size, which in turn explains the experimental signatures of $H_{\rm c2}$. These features stem from the non-BCS features due to the proximity to a Mott insulator. A high value of $H_{c2}$ near optimal doping is also sometimes associated to the presence of a quantum critical point in the literature \cite{Grissonnanche2014}. Our results do not depend on the presence of any quantum critical point near optimal doping. However, it will be an interesting future direction to connect our findings to a possible quantum critical point. Possible presence of competing orders can fine-tune the scenario by bringing in additional length scales. It should also be noted that our real space calculations naturally produce competing superconducting orders like extended s-wave order. However, the amplitudes of the extended s-wave order is extremely small and thus unlikely to have a significant effect on the LDOS. Our findings can have important implications on properties of other materials like ${\rm Fe}$-based superconductors and twisted bilayer graphene, where strong correlation physics is believed to play a crucial role~\cite{doi:10.1126/science.aal1575, PhysRevLett.112.177001, doi:10.1126/science.aao1797, Cao2018}.

\noindent{\it \bf Acknowledgement.} We thank M.~Randeria for valuable comments. We acknowledge support from Scheme of Promotion of Academic and Research collaboration (Grant No. 460). A.D. acknowledges support at Instituto de Ciencia de Materiales de Madrid - CSIC (under Grant No. PGC2018-097018-B-I00). The work of K.Y. was supported by the National Science Foundation (Grant No. DMR-1932796). The work of H.J.C. was supported by NSF-CAREER Grant No. DMR-2046570. The works of K.Y. and H.J.C. were performed at the National High Magnetic Field Laboratory, which is supported by National Science Foundation Cooperative Agreement No. DMR-1644779, and the State of Florida. The computations were facilitated by Dirac cluster at IISER Kolkata and Research Computing Cluster (RCC) at Florida State University.

\pagebreak
\widetext
\clearpage 
\normalsize
~\vspace{0.2cm} 
\setcounter{equation}{0}
\setcounter{figure}{0}
\setcounter{table}{0}
\setcounter{page}{1}
\makeatletter
\renewcommand{\theequation}{S\arabic{equation}}
\renewcommand{\thefigure}{S\arabic{figure}}
\renewcommand{\figurename}{FIG.}

\begin{center}
\textbf{\LARGE Supplementary material for `Enigma of the vortex state in a strongly correlated d-wave superconductor'}
\end{center}

\begin{center}
{\large Anushree Datta, Hitesh J. Changlani, Kun Yang, and Amit Ghosal}
\end{center}

\medskip

In this Supplementary Material (SM) we provide additional details of our method and results to further support the findings presented in the main text. We also describe the findings in the absence of strong correlation effects and compare them with the ones discussed in the main text.

{\section{Gutzwiller inhomogeneous mean-field theory}\label{sec:app1} 

The projection operators in Eq.~(1) of the main text prohibit formation of double occupancy on any lattice site. We employ Gutzwiller approximation (GA) method~\cite{KoNaveLee}, which captures the effects of the projection operators within Gutzwiller renormalization factors (GRFs)\cite{KoNaveLee,Zhang_1988}.
Within the GA, 
\begin{equation}
\langle \hat{c}_{i\sigma}\hat{c}_{j\sigma}\rangle_{\psi}\approx g^{t}_{ij}\langle \hat{c}_{i\sigma}\hat{c}_{j\sigma}\rangle_{\psi_{0}}~;~\langle \mathbf{\hat{S}}_{i}.\mathbf{\hat{S}}_{j}\rangle_{\psi}\approx g^{J}_{ij}\langle \mathbf{\hat{S}}_{i}.\mathbf{\hat{S}}_{j}\rangle_{\psi_{0}}~;~\langle \hat{n}_{i}\hat{n}_{j}\rangle_{\psi}\approx\langle \hat{n}_{i}\hat{n}_{j}\rangle_{\psi_{0}}~,
\end{equation}
where $\psi$ and $\psi_{0}$ represent the bases with and without the restriction on the doubly occupied states in the Hilbert space, respectively. Here, $g^{t}_{ij}$ and $g^{J}_{ij}$ are GRFs and they depend on the local densities $n_{i}$ as follows
\begin{equation}
g^{t}_{ij}=g^{t}_{i}g^{t}_{j}; ~g^{t}_{i}=\sqrt{\frac{1-n_{i}}{1-n_{i}/2}}~,
\label{GF1}
\end{equation}
\begin{equation}
g^{J}_{ij}=g^{J}_{i}g^{J}_{j} ; ~g^{J}_{i}=\frac{1}{1-n_{i}/2}~.
\label{GF3}
\end{equation}
The Gutzwiller augmented $t-J$ model then reads as
\begin{equation}
{\cal H}_{\rm {t-J}}=-\sum_{\langle ij \rangle, \sigma}tg^{t}_{ij}\left(e^{\phi_{ij}}\hat{c}^{\dagger}_{i\sigma}\hat{c}_{j\sigma}+{\rm H. c}\right)+ \sum_{\langle ij \rangle}J g^{J}_{ij} \left(\mathbf{\hat{S}}_{i}.\mathbf{\hat{S}}_{j}-\frac{\hat{n}_{i}\hat{n}_{j}}{4}\right)~.
\label{HGTJ}
\end{equation}
To obtain the mean-field Hamiltonian from Eq.~$(\ref{HGTJ})$, we minimize the functional $W=\langle\psi_{0}|{\cal H_{\rm t-J}}|\psi_{0}\rangle$ with respect to $|\psi_{0}\rangle$\cite{Yang_2009} under the constraints of having a fixed average electron density $\rho$ and a normalized wavefunction i.e., $\langle \psi_{0}|\psi_{0}\rangle=1$, during the minimization. This leads to~\cite{Yang_2009, Christensen_2011}
\begin{eqnarray}
{\cal H}_{\rm{MF}}&=&\sum_{\langle ij\rangle, \sigma} \frac{\partial W}{\partial \tau_{ij}} \left(\hat{c}^{\dagger}_{i\sigma}\hat{c}_{i\sigma}+ \rm H.c.\right)+ \sum_{i} \frac{\partial W}{\partial n_{i}}\hat{n}_{i}+ \sum_{\langle ij \rangle, \sigma} \frac{\partial W}{\partial \Delta_{ij}}\sigma \hat{c}_{i\sigma}\hat{c}_{j\overline\sigma}~,
\label{MF2}
\end{eqnarray}
where $\Delta_{ij}=\frac{1}{2}\sum_{\sigma}\langle\psi_{0}|\hat{c}_{i\sigma}\hat{c}_{j\overline{\sigma}}|\psi_{0}\rangle$ is the superconducting (SC) pairing amplitude, $n_{i}=\sum_{\sigma}\langle\psi_{0}|\hat{c}^{\dagger}_{i\sigma}\hat{c}_{i\sigma}|\psi_{0}\rangle$ is the local density, and $\tau_{ij}=\frac{1}{2}\sum_{\sigma}\langle\psi_{0}|\hat{c}^{\dagger}_{i\sigma}\hat{c}_{j\sigma}|\psi_{0}\rangle$ is the Fock shift. 
Calculating the derivatives in Eq.~(\ref{MF2}), we finally obtain,
\begin{eqnarray}
{\cal H}_{\rm{MF}}&=&\sum_{i\alpha\sigma}\left(-t-\frac {J}{4}\left(3g^{J}_{i\alpha}-1\right)\tau_{i\alpha}\right) e^{\phi^{\alpha}_{i}}\hat{c}^{\dagger}_{i\sigma}\hat{c}_{i+\alpha\sigma}\nonumber\\&+&\sum_{i\alpha} \left(-\frac {J}{4}\left(3g^{J}_{i\alpha}+1\right)\Delta_{i\alpha}\hat{c}^{\dagger}_{i\uparrow}\hat{c}^{\dagger}_{i+\alpha\downarrow}+\rm{H.c.}\right)\nonumber\\&+& \sum_{i}\left(-\mu+\mu_{i}\right)\hat{n}_{i}~.
\label{HMF}
\end{eqnarray}
The nearest neighbor bonds are denoted as $\alpha$ with nearest neighbor site of $i$ being $i+\alpha$ and $\mu_{i}$ is an effective local chemical potential, given by 
\begin{equation}
\mu_{i}=\mu^{g}_{i}-\frac{J}{4}\sum_{\alpha} n_{i+\alpha}~,
\label{mueff1}
\end{equation} 
where 
\begin{equation}
\mu^{g}_{i}=J\sum_{\alpha} \left( \frac{\tau_{i\alpha}\tau^{\ast}_{i\alpha}}{4}+ \frac{\Delta_{i\alpha}\Delta^{\ast}_{i\alpha}}{4}\right)\frac{dg^{J}_{ij}}{dn_{i}}-\sum_{\alpha}t\frac{dg^{t}_{i\alpha}}{dn_{i}}\left(e^{i\phi^{\alpha}_{i}}\tau_{i\alpha}+e^{-i\phi^{\alpha}_{i}}\tau^{\ast}_{i\alpha}\right)~. 
\label{mueff2}
\end{equation}
The derivatives of the Gutzwiller factors in $\mu^{g}_i$ are analytically calculated using the expressions in Eqs.~(\ref{GF1}, \ref{GF3}). 

As discussed in the model and methods of the main text, the usual inhomogeneous mean-field theory (IMT) calculations have no double occupancy restrictions. In IMT, the GRFs are set to be unity in Eq.~(\ref{HMF}). In IMT, we tune $J$ values for each doping in such a way that both IMT and GIMT yield the same d-wave gap when magnetic field is zero~\cite{Garg_2008}. The results from IMT will be presented in the next section for a contrast with the equivalent GIMT results in the main text.
\medskip

{\bf Bogoliubov-de Gennes equations.} ${\cal H}_{\rm MF}$ is diagonalized using the Bogoliubov-de Gennes (BdG) transformations $\hat{c}_{i\sigma}=\sum_{n}\left(\gamma_{n\sigma}u_{i,n}-\gamma^{\dagger}_{n\overline{\sigma}}v^{\ast}_{i,n}\right)$, where $\gamma_{n\sigma}$ and $\gamma^{\dagger}_{n\sigma}$ are the creation and annihilation operators of the Bogoliubov quasiparticles and $u_{i,n}$ and $v^{\ast}_{i,n}$ are the eigenfunctions with eigenvalues $E_n$. The resulting eigensystem is then given by
\begin{equation}
\begin{pmatrix}
\hat{\xi} & \hat{\Delta}\\
\hat{\Delta}^{*} & -\hat{\xi}^{*}
\end{pmatrix}
\begin{pmatrix}
u_{n}\\
v_{n}
\end{pmatrix}
= E_{n}
\begin{pmatrix}
u_{n}\\
v_{n}
\end{pmatrix}~,
\label{bdg}
\end{equation}
where 
\begin{equation}
\hat{\xi}u_{i, n}=-\sum_{\alpha} \left(-tg^{t}_{i\alpha}-\frac {J}{4}\left(3g^{J}_{i\alpha}-1\right)\right)e^{i\phi^{\alpha}_{i}}u_{i+\alpha, n}-\left(\mu-\mu_{i}\right)u_{i,n}~,
\label{bdg1}
\end{equation}
\begin{equation}
\hat{\Delta}v_{i,n}=-\frac {J}{4}\sum_{\alpha}\left(3g^{J}_{i\alpha}+1\right)\Delta_{i\alpha} v_{i+\alpha,n}~.
\label{bdg2}
\end{equation}
To improve the resolution of physical quantities, like the local density of states, we exploit the magnetic translation symmetry of the Hamiltonian in Eq.~(\ref{HMF}) and employ a repeated zone scheme (RZS) in the mean-field calculations~\cite{Ghosal_2002, Wang_1995}. 
We consider our system to be made up of a periodic array of identical magnetic unit cells (UCs) of size $N_{x}\times N_{y}$, enclosing two SC flux quanta, where $N_{x}=N_{y}/2$. Within the RZS, we solve the BdG matrix in one magnetic UC and expand the resulting wavefunctions in the entire system made up of $P\times Q$ UCs, using Bloch theorem. For most of the results, we take $N_x$, $N_y$, $P$, and $Q$ to be $24$, $48$, $16$, and $8$, respectively. Thus the total system, which follows periodic boundary condition, is composed of $256$ vortices. In Fig.~4(b) of the main text, we adjust 
the values of $N_x$, $N_y$, $P$, and $Q$ in such a way that the number of SC flux quanta per unit cell gives the magnetic field.

The RZS in the BdG equations is incorporated by using ideas behind the Bloch's theorem in our system which is made up of a periodic array of the magnetic UCs~\cite{Ghosal_2002, Wang_1995}. We consider a magnetic translational vector $\tau_{R}$, which translates a lattice vector $\mathbf{r}$ to $\mathbf{r+R}$ and commutes with the Hamiltonian ${\cal H}_{MF}$. Thus the eigenstates of the translational operator can be used to block diagonalize ${\cal H}_{MF}$. Here $\mathbf{R}$ denotes coordinates of the magnetic unit cells. We work with the Landau gauge $\mathbf{A(r)}=(0,Hx)$ and within this gauge the magnetic translation operators are $\langle \mathbf{r}|\tau_{R}| \mathbf{r}^{\prime}\rangle = \delta_{r, r+R} e^{-ibR_{x}(i_{y}+ R_{y})}$, where $i_y$ denotes $y$ coordinates of lattice sites in each of the magnetic unit cells. Following the Bloch's description on the vortex lattice of periodicity $R$, we then do the following tranformations of eigenvetors that block diagonalizes the BdG equations
\begin{equation}
u_{i,n}(\mathbf{R}) \rightarrow  e^{i\mathbf{k.R}} u^{k}_{i,n}e^{-ib\left(i_{y}+R_{y}\right)R_{x}},~\\
v_{i,n}(\mathbf{R}) \rightarrow e^{i\mathbf{k.R}} v^{k}_{i,n}e^{ib\left(i_{y}+R_{y}\right)R_{x}}~,
\end{equation}

where $\mathbf{k}$ is the Fourier transform of $\mathbf{R}$. The block diagonal BdG eigen-equations for each $\mathbf{k}$ then becomes

\begin{equation}
\begin{pmatrix}
\hat{\xi}(\mathbf{k}) & \hat{\Delta}(\mathbf{k})\\
\hat{\Delta}^{*}(\mathbf{k}) & -\hat{\xi}^{*}(\mathbf{k})
\end{pmatrix}
\begin{pmatrix}
u^{k}_{n}\\
v^{k}_{n}
\end{pmatrix}
= E_{k,n}
\begin{pmatrix}
u^{k}_{n}\\
v^{k}_{n}
\end{pmatrix}~,
\label{bdg}
\end{equation}
where the operators in the bulk of the UCs are given by 
\begin{equation}
\hat{\xi}(\mathbf{k})u^{k}_{i, n}=-\sum_{\alpha} \left(-tg^{t}_{i\alpha}-\frac {J}{4}\left(3g^{J}_{i\alpha}-1\right)\tau_{i\alpha}\right)e^{i\phi^{\alpha}_{i}}u^{k}_{i+\alpha, n}-\left(\mu-\mu_{i}\right)u^{k}_{i,n}~,
\label{rzs1}
\end{equation}
\begin{equation}
\hat{\Delta}(\mathbf{k})v^{k}_{i, n}=-\frac {J}{4}\sum_{\alpha}\left(3g^{J}_{i\alpha}+1\right)\Delta_{i\alpha} v^{k}_{i+\alpha, n}~.
\label{rzs2}
\end{equation}
In the RZS, the self-consistent equations in the bulk of the UCs are then given by
\begin{equation}
\Delta_{i\alpha}=\sum_{n,q}\frac{1}{N_{e}} \Big[u^{q}_{i, n}v^{q\ast}_{i+\alpha,n}f(-E_{qn})+ u^{q}_{i+\alpha,n}v^{q,\ast}_{i,n}f(E_{qn})\Big]~,
\label{rzsop1}
\end{equation}
\begin{equation}
\tau_{i\alpha}=\sum_{n,q}\frac{1}{N_{e}}\Big[u^{q,\ast}_{i,n}u^{q}_{i+\alpha,n}f(E_{q n})+v^{q}_{i,n}v^{q,\ast}_{i+\alpha,n}f(-E_{qn})\Big]~,
\label{rzsop2}
\end{equation}
\begin{equation}
n_{i}=2\sum_{n,q}\frac{1}{N_{e}}\Big[u^{q,\ast}_{i,n}u^{q}_{i,n}f(E_{qn})+v^{q,\ast}_{i,n}v^{q}_{i,n}f(-E_{qn})\Big]~,
\label{rzsop3}
\end{equation}
where $N_{e}$ is the total number of eigenstates, given by $N_{e}=N_{x}P\times N_{y}Q$. At the boundaries of the unit cells the above equations will be suitably modified taking care of the phase factors arising out of Bloch functions, as well as from the Peierls factor due to the orbital magnetic field. The resulting eigenvalue problem is then solved for all local self-consistent parameters $\Delta_{i\alpha}$, $\tau_{i\alpha}$, and $n_{i}$. 

\begin{figure}
\centering
    \includegraphics[width=0.55\textwidth]{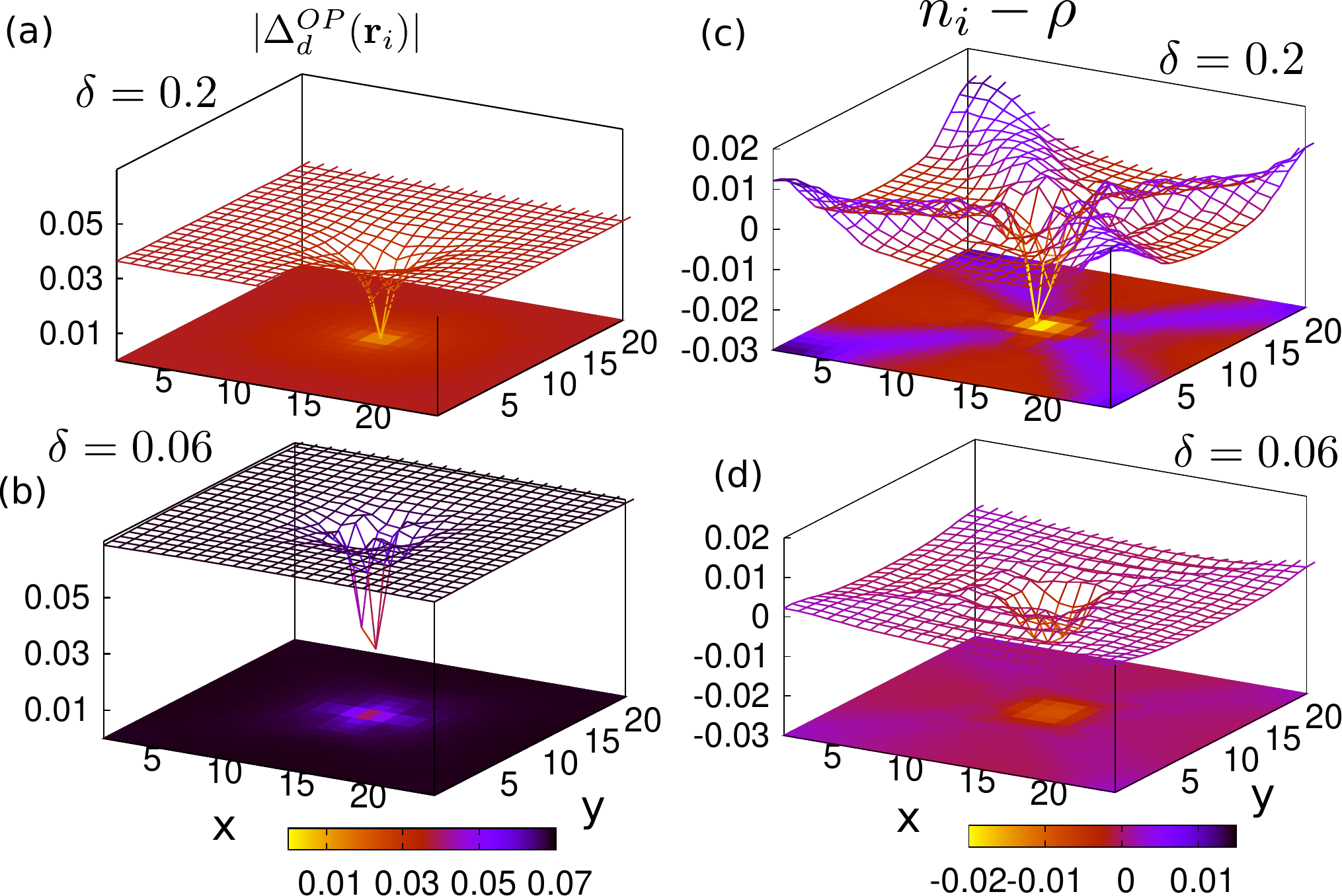} 
  \caption{d-wave SC $|\Delta^{\rm OP}_{\rm d}({\bf r}_{i})|$ profiles and local electronic density variation with respect to the average values $n(i)-\rho$ on a magnetic unit cell of size $24\times24$, for doping values $\delta=0.2$ (a, c) and $0.06$ (b, d), obtained from IMT calculations. }
\label{fig:figIMTOP}
\end{figure}

\section{Order parameters and local density of states from IMT calculations}
In this section of the SM, we present the SC order parameter, electron density, and local density of states (LDOS) obtained within IMT calculation where the effects of the strong correlations are ignored by putting the GRFs to unity. We further contrast IMT results with GIMT results presented in the main manuscript and show that the important finding of the crossover in the nature of the vortex core from metallic to Mott insulating is solely due to strong correlations.
 
 {\bf Order parameters.}~In Fig.~\ref{fig:figIMTOP} we first show the spatial map of d-wave SC order parameter $|\Delta^{\rm OP}_{\rm d}|$ around a vortex core obtained from the IMT calculation for optimal doping $\delta=0.2$ (a) and heavily underdoping $\delta=0.06$ (b). The exchange parameters for IMT with $\delta=0.2$ and $\delta=0.06$ are fixed at $J=1.2$ and $1.5$ respectively, in order to match the zero field d-wave SC gap with their GIMT counterparts, for a reasonable comparison. As seen in (a) and (b), $|\Delta^{\rm OP}_{\rm d}|$ follows a conical fall to zero at the vortex core for both $\delta=0.2$ and $\delta=0.06$. While the conical fall for $\delta=0.2$ is also observed in GIMT, shown in Fig.~1(b) of the main text, the ``flat-bottom bowl" shape of $|\Delta^{\rm OP}_{\rm d}|$ seen in Fig.~1(d) of the main text in GIMT is absent in IMT for heavily underdoping $\delta=0.06$. Thus, the IMT results clearly show that the ``flat-bottom bowl" shape of $|\Delta^{\rm OP}_{\rm d}|$ obtained in GIMT is solely due to the effect of strong correlations. Another noticeable feature in (a) and (b) is the decrease of the core size with decrease in doping. This feature is also shown in the Fig.~4(a) of the main text for more doping values.  

In order to compare the electron density of the vortex core, we plot in Fig.~\ref{fig:figIMTOP} (c) and (d) the spatial map of the electron density relative to the density away from the vortex core, $n_{i}-\rho$ within IMT calculation. In contrast to our findings of GIMT in Fig.~2 of the main text, density at the vortex core $n_{{\bf r}_v}$ is lesser than $\rho$ for both optimal doping and heavily underdoping. As a result, $n_{{\bf r}_v}$ remains far from half-filling. 
\begin{figure}
\centering
    \includegraphics[width=0.65\textwidth]{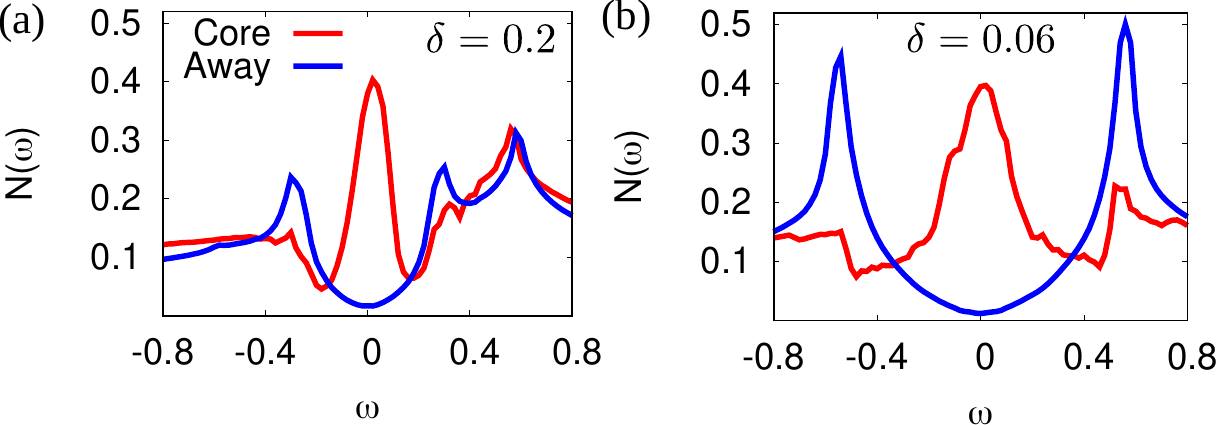} 
  \caption{LDOS profiles at the vortex core (red traces) and away from the core (blue traces) for (a) $\delta=0.2$ and (b) $\delta=0.06$, calculated within the IMT calculation. A prominent mid-gap peak persists at both doping values.}
\label{fig:figIMTDOS}
\end{figure}

{\bf Local density of states.} To illustrate the physical effect of the differences in the order parameter profiles in IMT and GIMT, we now show in Fig.~\ref{fig:figIMTDOS} LDOS profiles obtained from IMT calculations at $\delta=0.2$ and $\delta=0.06$. At both dopings, the LDOS carry a broad peak of zero-energy core states at vortex cores, signaling their metallic nature. Hence, we again show that the lack of the zero-energy core states at vortex cores within GIMT in Fig~3 of the main text is solely due to the presence of strong correlations.

{\section{Role of GRFs on emerging Mottness at the vortex core in GIMT} \label{sec:app2}
We find that the effects of the local GRFs in the effective chemical potential play a key role in rendering Mottness at the vortex core. The effective chemical potential $\mu_{i}$ in GIMT as defined in Eq.~(\ref{mueff1}) and (\ref{mueff2}) includes local derivatives of the GRFs in $\mu^{g}_{i}$. 
\begin{figure}
\centering
    \includegraphics[width=0.55\textwidth]{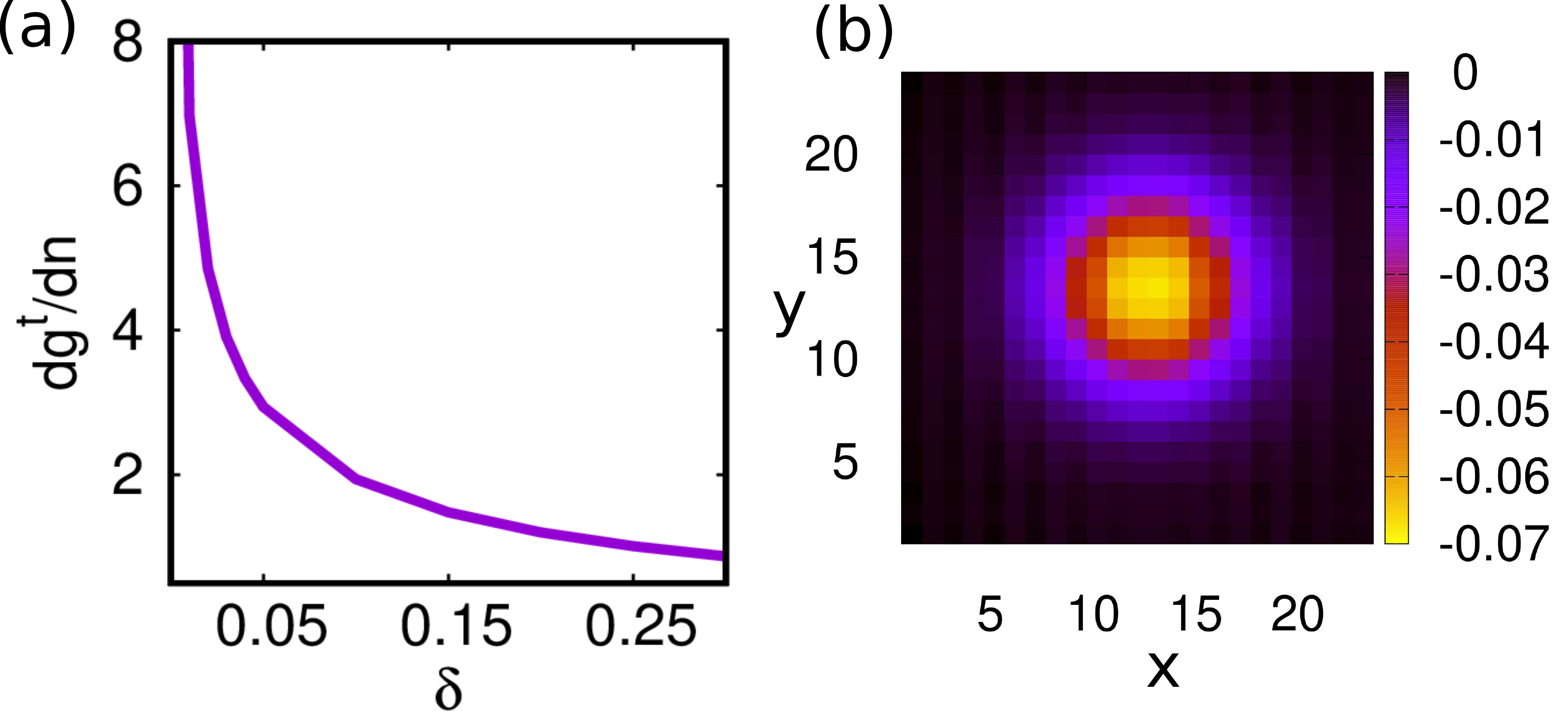}  
  \caption{(a) Variation of $dg^{t}/dn$ as a function of doping. The magnitude of the derivative rises sharply with decreasing doping. (b) Spatial profile of the variation of the effective chemical potential $\mu_{i}-\mu(\rm away)$ around a vortex core, from GIMT calculations at $\delta=0.06$. $\mu(\rm away)$ is the effective chemical potential away from the vortex core. $\mu_{i}$ carry a prominent dip at the vortex core due to the local effects of the GRFs. }
\label{fig:fig7}
\end{figure}
As shown in Fig.~\ref{fig:fig7}(a), the magnitude of the derivative of the GRF $dg^{t}/dn$ rises significantly when the doping approaches half-filling in the homogeneous case. This causes $\mu^{g}_{i}$ to play a key role in creating a substantial chemical potential difference between the normal state at the vortex core region and the superconducting region, and hence dictating the density landscape in the underdoped region. Apart from the $\mu^{g}_{i}$ term, the chemical potential difference includes the difference due to the condensation energy in the superconducting region \cite{Khomskii_1995,Blatter_1996} and the contribution from the Hartree shift. We checked by setting $\mu^{g}=0$, that these other contributions are not sufficient to make the density at the vortex core reach half-filling. Fig.~\ref{fig:fig7}(b) features the profile of ${\mu}_{i}-{\mu}({\rm away})$ over a $24\times24$ magnetic cell around a vortex, at $\delta=0.06$. Here ${\mu}({\rm away})$ is the effective chemical potential away from the vortex core which is close to the uniform value resulting in the filling $\rho$. The dominance of the derivatives of the GRFs at $\delta=0.06$ causes $\mu_{i}$ to self-consistently build a prominent dip in the local potential around the vortex core attracting electrons to locally pile up forming a region in the vortex core where local density approaches half-filling, and the pairing amplitude remains depleted. Such vortex cores are found energetically favorable at underdoping.


\begin{thebibliography}{52}%
\makeatletter
\providecommand \@ifxundefined [1]{%
 \@ifx{#1\undefined}
}%
\providecommand \@ifnum [1]{%
 \ifnum #1\expandafter \@firstoftwo
 \else \expandafter \@secondoftwo
 \fi
}%
\providecommand \@ifx [1]{%
 \ifx #1\expandafter \@firstoftwo
 \else \expandafter \@secondoftwo
 \fi
}%
\providecommand \natexlab [1]{#1}%
\providecommand \enquote  [1]{``#1''}%
\providecommand \bibnamefont  [1]{#1}%
\providecommand \bibfnamefont [1]{#1}%
\providecommand \citenamefont [1]{#1}%
\providecommand \href@noop [0]{\@secondoftwo}%
\providecommand \href [0]{\begingroup \@sanitize@url \@href}%
\providecommand \@href[1]{\@@startlink{#1}\@@href}%
\providecommand \@@href[1]{\endgroup#1\@@endlink}%
\providecommand \@sanitize@url [0]{\catcode `\\12\catcode `\$12\catcode
  `\&12\catcode `\#12\catcode `\^12\catcode `\_12\catcode `\%12\relax}%
\providecommand \@@startlink[1]{}%
\providecommand \@@endlink[0]{}%
\providecommand \url  [0]{\begingroup\@sanitize@url \@url }%
\providecommand \@url [1]{\endgroup\@href {#1}{\urlprefix }}%
\providecommand \urlprefix  [0]{URL }%
\providecommand \Eprint [0]{\href }%
\providecommand \doibase [0]{http://dx.doi.org/}%
\providecommand \selectlanguage [0]{\@gobble}%
\providecommand \bibinfo  [0]{\@secondoftwo}%
\providecommand \bibfield  [0]{\@secondoftwo}%
\providecommand \translation [1]{[#1]}%
\providecommand \BibitemOpen [0]{}%
\providecommand \bibitemStop [0]{}%
\providecommand \bibitemNoStop [0]{.\EOS\space}%
\providecommand \EOS [0]{\spacefactor3000\relax}%
\providecommand \BibitemShut  [1]{\csname bibitem#1\endcsname}%
\let\auto@bib@innerbib\@empty
\bibitem [{\citenamefont {Kosterlitz}\ and\ \citenamefont
  {Thouless}(1973)}]{Kosterlitz_1973}%
  \BibitemOpen
  \bibfield  {author} {\bibinfo {author} {\bibfnamefont {J.~M.}\ \bibnamefont
  {Kosterlitz}}\ and\ \bibinfo {author} {\bibfnamefont {D.~J.}\ \bibnamefont
  {Thouless}},\ }\href {\doibase 10.1088/0022-3719/6/7/010} {\bibfield
  {journal} {\bibinfo  {journal} {J. Phys. C: Solid St. Phys.}\
  }\textbf {\bibinfo {volume} {6}},\ \bibinfo {pages} {1181} (\bibinfo {year}
  {1973})}\BibitemShut {NoStop}%
\bibitem [{\citenamefont {Kosterlitz}(2017)}]{RevModPhys.89.040501}%
  \BibitemOpen
  \bibfield  {author} {\bibinfo {author} {\bibfnamefont {J.~M.}\ \bibnamefont
  {Kosterlitz}},\ }\href {\doibase 10.1103/RevModPhys.89.040501} {\bibfield
  {journal} {\bibinfo  {journal} {Rev. Mod. Phys.}\ }\textbf {\bibinfo {volume}
  {89}},\ \bibinfo {pages} {040501} (\bibinfo {year} {2017})}\BibitemShut
  {NoStop}%
\bibitem [{\citenamefont {Abrikosov}(1957)}]{AVL}%
  \BibitemOpen
  \bibfield  {author} {\bibinfo {author} {\bibfnamefont {A.}~\bibnamefont
  {Abrikosov}},\ }\href {\doibase https://doi.org/10.1016/0022-3697(57)90083-5}
  {\bibfield  {journal} {\bibinfo  {journal} {J. Phys. Chem. Solids}\ }\textbf
  {\bibinfo {volume} {2}},\ \bibinfo {pages} {199 } (\bibinfo {year}
  {1957})}\BibitemShut {NoStop}%
\bibitem [{\citenamefont {Kopnin}(2001)}]{Book2}%
  \BibitemOpen
  \bibfield  {author} {\bibinfo {author} {\bibfnamefont {N.}~\bibnamefont
  {Kopnin}},\ }\href@noop {} {\emph {\bibinfo {title} {{Vortices in Type-II
  Superconductors: Structure and Dynamics}}}}\ (\bibinfo  {publisher} {Oxford
  University Press},\ \bibinfo {year} {2001})\BibitemShut {NoStop}%
\bibitem [{\citenamefont {Tinkham}(2004)}]{Tinkham}%
  \BibitemOpen
  \bibfield  {author} {\bibinfo {author} {\bibfnamefont {M.}~\bibnamefont
  {Tinkham}},\ }\href {https://books.google.es/books?id=VpUk3NfwDIkC} {\emph
  {\bibinfo {title} {Introduction to Superconductivity}}},\ Dover Books on
  Physics Series\ (\bibinfo  {publisher} {Dover Publications},\ \bibinfo {year}
  {2004})\BibitemShut {NoStop}%
\bibitem [{\citenamefont {Yin}\ \emph {et~al.}(2015)\citenamefont {Yin},
  \citenamefont {Wu}, \citenamefont {Wang}, \citenamefont {Ye}, \citenamefont
  {Gong}, \citenamefont {Hou}, \citenamefont {Shan}, \citenamefont {Li},
  \citenamefont {Liang}, \citenamefont {Wu}, \citenamefont {Li}, \citenamefont
  {Ting}, \citenamefont {Wang}, \citenamefont {Hu}, \citenamefont {Hor},
  \citenamefont {Ding},\ and\ \citenamefont {Pan}}]{Yin2015}%
  \BibitemOpen
  \bibfield  {author} {\bibinfo {author} {\bibfnamefont {J.-X.}\ \bibnamefont
  {Yin}}, \bibinfo {author} {\bibfnamefont {Z.}~\bibnamefont {Wu}}, \bibinfo
  {author} {\bibfnamefont {J.-H.}\ \bibnamefont {Wang}}, \bibinfo {author}
  {\bibfnamefont {Z.-Y.}\ \bibnamefont {Ye}}, \bibinfo {author} {\bibfnamefont
  {J.}~\bibnamefont {Gong}}, \bibinfo {author} {\bibfnamefont {X.-Y.}\
  \bibnamefont {Hou}}, \bibinfo {author} {\bibfnamefont {L.}~\bibnamefont
  {Shan}}, \bibinfo {author} {\bibfnamefont {A.}~\bibnamefont {Li}}, \bibinfo
  {author} {\bibfnamefont {X.-J.}\ \bibnamefont {Liang}}, \bibinfo {author}
  {\bibfnamefont {X.-X.}\ \bibnamefont {Wu}}, \bibinfo {author} {\bibfnamefont
  {J.}~\bibnamefont {Li}}, \bibinfo {author} {\bibfnamefont {C.-S.}\
  \bibnamefont {Ting}}, \bibinfo {author} {\bibfnamefont {Z.-Q.}\ \bibnamefont
  {Wang}}, \bibinfo {author} {\bibfnamefont {J.-P.}\ \bibnamefont {Hu}},
  \bibinfo {author} {\bibfnamefont {P.-H.}\ \bibnamefont {Hor}}, \bibinfo
  {author} {\bibfnamefont {H.}~\bibnamefont {Ding}}, \ and\ \bibinfo {author}
  {\bibfnamefont {S.~H.}\ \bibnamefont {Pan}},\ }\href {\doibase
  10.1038/nphys3371} {\bibfield  {journal} {\bibinfo  {journal} {Nature
  Physics}\ }\textbf {\bibinfo {volume} {11}},\ \bibinfo {pages} {543}
  (\bibinfo {year} {2015})}\BibitemShut {NoStop}%
\bibitem [{\citenamefont {Gazdi\ifmmode~\acute{c}\else \'{c}\fi{}}\ \emph
  {et~al.}(2021)\citenamefont {Gazdi\ifmmode~\acute{c}\else \'{c}\fi{}},
  \citenamefont {Maggio-Aprile}, \citenamefont {Gu},\ and\ \citenamefont
  {Renner}}]{PhysRevX.11.031040}%
  \BibitemOpen
  \bibfield  {author} {\bibinfo {author} {\bibfnamefont {T.}~\bibnamefont
  {Gazdi\ifmmode~\acute{c}\else \'{c}\fi{}}}, \bibinfo {author} {\bibfnamefont
  {I.}~\bibnamefont {Maggio-Aprile}}, \bibinfo {author} {\bibfnamefont
  {G.}~\bibnamefont {Gu}}, \ and\ \bibinfo {author} {\bibfnamefont
  {C.}~\bibnamefont {Renner}},\ }\href {\doibase 10.1103/PhysRevX.11.031040}
  {\bibfield  {journal} {\bibinfo  {journal} {Phys. Rev. X}\ }\textbf {\bibinfo
  {volume} {11}},\ \bibinfo {pages} {031040} (\bibinfo {year}
  {2021})}\BibitemShut {NoStop}%
\bibitem [{\citenamefont {Edkins}\ \emph {et~al.}(2019)\citenamefont {Edkins},
  \citenamefont {Kostin}, \citenamefont {Fujita}, \citenamefont {Mackenzie},
  \citenamefont {Eisaki}, \citenamefont {Uchida}, \citenamefont {Sachdev},
  \citenamefont {Lawler}, \citenamefont {Kim}, \citenamefont {Davis},\ and\
  \citenamefont {Hamidian}}]{doi:10.1126/science.aat1773}%
  \BibitemOpen
  \bibfield  {author} {\bibinfo {author} {\bibfnamefont {S.~D.}\ \bibnamefont
  {Edkins}}, \bibinfo {author} {\bibfnamefont {A.}~\bibnamefont {Kostin}},
  \bibinfo {author} {\bibfnamefont {K.}~\bibnamefont {Fujita}}, \bibinfo
  {author} {\bibfnamefont {A.~P.}\ \bibnamefont {Mackenzie}}, \bibinfo {author}
  {\bibfnamefont {H.}~\bibnamefont {Eisaki}}, \bibinfo {author} {\bibfnamefont
  {S.}~\bibnamefont {Uchida}}, \bibinfo {author} {\bibfnamefont
  {S.}~\bibnamefont {Sachdev}}, \bibinfo {author} {\bibfnamefont {M.~J.}\
  \bibnamefont {Lawler}}, \bibinfo {author} {\bibfnamefont {E.-A.}\
  \bibnamefont {Kim}}, \bibinfo {author} {\bibfnamefont {J.~C.~S.}\
  \bibnamefont {Davis}}, \ and\ \bibinfo {author} {\bibfnamefont {M.~H.}\
  \bibnamefont {Hamidian}},\ }\href {\doibase 10.1126/science.aat1773}
  {\bibfield  {journal} {\bibinfo  {journal} {Science}\ }\textbf {\bibinfo
  {volume} {364}},\ \bibinfo {pages} {976} (\bibinfo {year}
  {2019})}\BibitemShut {NoStop}%
\bibitem [{\citenamefont {Maggio-Aprile}\ \emph {et~al.}(1995)\citenamefont
  {Maggio-Aprile}, \citenamefont {Renner}, \citenamefont {Erb}, \citenamefont
  {Walker},\ and\ \citenamefont {Fischer}}]{PhysRevLett.75.2754}%
  \BibitemOpen
  \bibfield  {author} {\bibinfo {author} {\bibfnamefont {I.}~\bibnamefont
  {Maggio-Aprile}}, \bibinfo {author} {\bibfnamefont {C.}~\bibnamefont
  {Renner}}, \bibinfo {author} {\bibfnamefont {A.}~\bibnamefont {Erb}},
  \bibinfo {author} {\bibfnamefont {E.}~\bibnamefont {Walker}}, \ and\ \bibinfo
  {author} {\bibfnamefont {O.}~\bibnamefont {Fischer}},\ }\href {\doibase
  10.1103/PhysRevLett.75.2754} {\bibfield  {journal} {\bibinfo  {journal}
  {Phys. Rev. Lett.}\ }\textbf {\bibinfo {volume} {75}},\ \bibinfo {pages}
  {2754} (\bibinfo {year} {1995})}\BibitemShut {NoStop}%
\bibitem [{\citenamefont {Pan}\ \emph {et~al.}(2000)\citenamefont {Pan},
  \citenamefont {Hudson}, \citenamefont {Gupta}, \citenamefont {Ng},
  \citenamefont {Eisaki}, \citenamefont {Uchida},\ and\ \citenamefont
  {Davis}}]{PhysRevLett.85.1536}%
  \BibitemOpen
  \bibfield  {author} {\bibinfo {author} {\bibfnamefont {S.~H.}\ \bibnamefont
  {Pan}}, \bibinfo {author} {\bibfnamefont {E.~W.}\ \bibnamefont {Hudson}},
  \bibinfo {author} {\bibfnamefont {A.~K.}\ \bibnamefont {Gupta}}, \bibinfo
  {author} {\bibfnamefont {K.-W.}\ \bibnamefont {Ng}}, \bibinfo {author}
  {\bibfnamefont {H.}~\bibnamefont {Eisaki}}, \bibinfo {author} {\bibfnamefont
  {S.}~\bibnamefont {Uchida}}, \ and\ \bibinfo {author} {\bibfnamefont {J.~C.}\
  \bibnamefont {Davis}},\ }\href {\doibase 10.1103/PhysRevLett.85.1536}
  {\bibfield  {journal} {\bibinfo  {journal} {Phys. Rev. Lett.}\ }\textbf
  {\bibinfo {volume} {85}},\ \bibinfo {pages} {1536} (\bibinfo {year}
  {2000})}\BibitemShut {NoStop}%
\bibitem [{\citenamefont {Wang}\ and\ \citenamefont
  {MacDonald}(1995)}]{PhysRevB.52.R3876}%
  \BibitemOpen
  \bibfield  {author} {\bibinfo {author} {\bibfnamefont {Y.}~\bibnamefont
  {Wang}}\ and\ \bibinfo {author} {\bibfnamefont {A.~H.}\ \bibnamefont
  {MacDonald}},\ }\href {\doibase 10.1103/PhysRevB.52.R3876} {\bibfield
  {journal} {\bibinfo  {journal} {Phys. Rev. B}\ }\textbf {\bibinfo {volume}
  {52}},\ \bibinfo {pages} {R3876} (\bibinfo {year} {1995})}\BibitemShut
  {NoStop}%
\bibitem [{\citenamefont {Nikoli\ifmmode~\acute{c}\else \'{c}\fi{}}\ \emph
  {et~al.}(2006)\citenamefont {Nikoli\ifmmode~\acute{c}\else \'{c}\fi{}},
  \citenamefont {Sachdev},\ and\ \citenamefont
  {Bartosch}}]{PhysRevB.74.144516}%
  \BibitemOpen
  \bibfield  {author} {\bibinfo {author} {\bibfnamefont {P.}~\bibnamefont
  {Nikoli\ifmmode~\acute{c}\else \'{c}\fi{}}}, \bibinfo {author} {\bibfnamefont
  {S.}~\bibnamefont {Sachdev}}, \ and\ \bibinfo {author} {\bibfnamefont
  {L.}~\bibnamefont {Bartosch}},\ }\href {\doibase 10.1103/PhysRevB.74.144516}
  {\bibfield  {journal} {\bibinfo  {journal} {Phys. Rev. B}\ }\textbf {\bibinfo
  {volume} {74}},\ \bibinfo {pages} {144516} (\bibinfo {year}
  {2006})}\BibitemShut {NoStop}%
\bibitem [{\citenamefont {Melikyan}\ and\ \citenamefont {Te\ifmmode
  \check{s}\else \v{s}\fi{}anovi\ifmmode~\acute{c}\else
  \'{c}\fi{}}(2007)}]{PhysRevB.76.094509}%
  \BibitemOpen
  \bibfield  {author} {\bibinfo {author} {\bibfnamefont {A.}~\bibnamefont
  {Melikyan}}\ and\ \bibinfo {author} {\bibfnamefont {Z.}~\bibnamefont
  {Te\ifmmode \check{s}\else \v{s}\fi{}anovi\ifmmode~\acute{c}\else
  \'{c}\fi{}}},\ }\href {\doibase 10.1103/PhysRevB.76.094509} {\bibfield
  {journal} {\bibinfo  {journal} {Phys. Rev. B}\ }\textbf {\bibinfo {volume}
  {76}},\ \bibinfo {pages} {094509} (\bibinfo {year} {2007})}\BibitemShut
  {NoStop}%
\bibitem [{\citenamefont {Tsuchiura}\ \emph {et~al.}(2003)\citenamefont
  {Tsuchiura}, \citenamefont {Ogata}, \citenamefont {Tanaka},\ and\
  \citenamefont {Kashiwaya}}]{PhysRevB.68.012509}%
  \BibitemOpen
  \bibfield  {author} {\bibinfo {author} {\bibfnamefont {H.}~\bibnamefont
  {Tsuchiura}}, \bibinfo {author} {\bibfnamefont {M.}~\bibnamefont {Ogata}},
  \bibinfo {author} {\bibfnamefont {Y.}~\bibnamefont {Tanaka}}, \ and\ \bibinfo
  {author} {\bibfnamefont {S.}~\bibnamefont {Kashiwaya}},\ }\href {\doibase
  10.1103/PhysRevB.68.012509} {\bibfield  {journal} {\bibinfo  {journal} {Phys.
  Rev. B}\ }\textbf {\bibinfo {volume} {68}},\ \bibinfo {pages} {012509}
  (\bibinfo {year} {2003})}\BibitemShut {NoStop}%
\bibitem [{\citenamefont {Vafek}\ \emph {et~al.}(2001)\citenamefont {Vafek},
  \citenamefont {Melikyan}, \citenamefont {Franz},\ and\ \citenamefont
  {Te\ifmmode \check{s}\else \v{s}\fi{}anovi\ifmmode~\acute{c}\else
  \'{c}\fi{}}}]{PhysRevB.63.134509}%
  \BibitemOpen
  \bibfield  {author} {\bibinfo {author} {\bibfnamefont {O.}~\bibnamefont
  {Vafek}}, \bibinfo {author} {\bibfnamefont {A.}~\bibnamefont {Melikyan}},
  \bibinfo {author} {\bibfnamefont {M.}~\bibnamefont {Franz}}, \ and\ \bibinfo
  {author} {\bibfnamefont {Z.}~\bibnamefont {Te\ifmmode \check{s}\else
  \v{s}\fi{}anovi\ifmmode~\acute{c}\else \'{c}\fi{}}},\ }\href {\doibase
  10.1103/PhysRevB.63.134509} {\bibfield  {journal} {\bibinfo  {journal} {Phys.
  Rev. B}\ }\textbf {\bibinfo {volume} {63}},\ \bibinfo {pages} {134509}
  (\bibinfo {year} {2001})}\BibitemShut {NoStop}%
\bibitem [{\citenamefont {Zhu}\ and\ \citenamefont
  {Ting}(2001)}]{PhysRevLett.87.147002}%
  \BibitemOpen
  \bibfield  {author} {\bibinfo {author} {\bibfnamefont {J.-X.}\ \bibnamefont
  {Zhu}}\ and\ \bibinfo {author} {\bibfnamefont {C.~S.}\ \bibnamefont {Ting}},\
  }\href {\doibase 10.1103/PhysRevLett.87.147002} {\bibfield  {journal}
  {\bibinfo  {journal} {Phys. Rev. Lett.}\ }\textbf {\bibinfo {volume} {87}},\
  \bibinfo {pages} {147002} (\bibinfo {year} {2001})}\BibitemShut {NoStop}%
\bibitem [{\citenamefont {Ghosal}\ \emph {et~al.}(2002)\citenamefont {Ghosal},
  \citenamefont {Kallin},\ and\ \citenamefont
  {Berlinsky}}]{PhysRevB.66.214502}%
  \BibitemOpen
  \bibfield  {author} {\bibinfo {author} {\bibfnamefont {A.}~\bibnamefont
  {Ghosal}}, \bibinfo {author} {\bibfnamefont {C.}~\bibnamefont {Kallin}}, \
  and\ \bibinfo {author} {\bibfnamefont {A.~J.}\ \bibnamefont {Berlinsky}},\
  }\href {\doibase 10.1103/PhysRevB.66.214502} {\bibfield  {journal} {\bibinfo
  {journal} {Phys. Rev. B}\ }\textbf {\bibinfo {volume} {66}},\ \bibinfo
  {pages} {214502} (\bibinfo {year} {2002})}\BibitemShut {NoStop}%
\bibitem [{\citenamefont {Zhang}\ \emph {et~al.}(2002)\citenamefont {Zhang},
  \citenamefont {Demler},\ and\ \citenamefont
  {Sachdev}}]{PhysRevB.66.094501}%
  \BibitemOpen
  \bibfield  {author} {\bibinfo {author} {\bibfnamefont {Y.}~\bibnamefont
  {Zhang}}, \bibinfo {author} {\bibfnamefont {E.}~\bibnamefont {Demler}}, \
  and\ \bibinfo {author} {\bibfnamefont {S.}\ \bibnamefont {Sachdev}},\
  }\href {\doibase 10.1103/PhysRevB.66.094501} {\bibfield  {journal} {\bibinfo
  {journal} {Phys. Rev. B}\ }\textbf {\bibinfo {volume} {66}},\ \bibinfo
  {pages} {094501} (\bibinfo {year} {2002})}\BibitemShut {NoStop}%
\bibitem [{\citenamefont {Himeda}\ \emph {et~al.}(1997)\citenamefont {Himeda},
  \citenamefont {Ogata}, \citenamefont {Tanaka},\ and\ \citenamefont
  {Kashiwaya}}]{doi:10.1143/JPSJ.66.3367}%
  \BibitemOpen
  \bibfield  {author} {\bibinfo {author} {\bibfnamefont {A.}~\bibnamefont
  {Himeda}}, \bibinfo {author} {\bibfnamefont {M.}~\bibnamefont {Ogata}},
  \bibinfo {author} {\bibfnamefont {Y.}~\bibnamefont {Tanaka}}, \ and\ \bibinfo
  {author} {\bibfnamefont {S.}~\bibnamefont {Kashiwaya}},\ }\href {\doibase
  10.1143/JPSJ.66.3367} {\bibfield  {journal} {\bibinfo  {journal} {Journal of
  the Physical Society of Japan}\ }\textbf {\bibinfo {volume} {66}},\ \bibinfo
  {pages} {3367} (\bibinfo {year} {1997})}\BibitemShut {NoStop}%
\bibitem [{\citenamefont {Ma\ifmmode~\acute{s}\else \'{s}\fi{}ka}\ and\
  \citenamefont {Mierzejewski}(2003)}]{PhysRevB.68.024513}%
  \BibitemOpen
  \bibfield  {author} {\bibinfo {author} {\bibfnamefont {M.~M.}\ \bibnamefont
  {Ma\ifmmode~\acute{s}\else \'{s}\fi{}ka}}\ and\ \bibinfo {author}
  {\bibfnamefont {M.}~\bibnamefont {Mierzejewski}},\ }\href {\doibase
  10.1103/PhysRevB.68.024513} {\bibfield  {journal} {\bibinfo  {journal} {Phys.
  Rev. B}\ }\textbf {\bibinfo {volume} {68}},\ \bibinfo {pages} {024513}
  (\bibinfo {year} {2003})}\BibitemShut {NoStop}%
\bibitem [{\citenamefont {Seo}\ \emph {et~al.}(2007)\citenamefont {Seo},
  \citenamefont {Chen},\ and\ \citenamefont {Hu}}]{PhysRevB.76.020511}%
  \BibitemOpen
  \bibfield  {author} {\bibinfo {author} {\bibfnamefont {K.}~\bibnamefont
  {Seo}}, \bibinfo {author} {\bibfnamefont {H.-D.}\ \bibnamefont {Chen}}, \
  and\ \bibinfo {author} {\bibfnamefont {J.}~\bibnamefont {Hu}},\ }\href
  {\doibase 10.1103/PhysRevB.76.020511} {\bibfield  {journal} {\bibinfo
  {journal} {Phys. Rev. B}\ }\textbf {\bibinfo {volume} {76}},\ \bibinfo
  {pages} {020511} (\bibinfo {year} {2007})}\BibitemShut {NoStop}%
\bibitem [{\citenamefont {Dai}\ \emph {et~al.}(2018)\citenamefont {Dai},
  \citenamefont {Zhang}, \citenamefont {Senthil},\ and\ \citenamefont
  {Lee}}]{PhysRevB.97.174511}%
  \BibitemOpen
  \bibfield  {author} {\bibinfo {author} {\bibfnamefont {Z.}~\bibnamefont
  {Dai}}, \bibinfo {author} {\bibfnamefont {Y.-H.}\ \bibnamefont {Zhang}},
  \bibinfo {author} {\bibfnamefont {T.}~\bibnamefont {Senthil}}, \ and\
  \bibinfo {author} {\bibfnamefont {P.~A.}\ \bibnamefont {Lee}},\ }\href
  {\doibase 10.1103/PhysRevB.97.174511} {\bibfield  {journal} {\bibinfo
  {journal} {Phys. Rev. B}\ }\textbf {\bibinfo {volume} {97}},\ \bibinfo
  {pages} {174511} (\bibinfo {year} {2018})}\BibitemShut {NoStop}%
\bibitem [{\citenamefont {Bru{\'e}r}\ \emph {et~al.}(2016)\citenamefont
  {Bru{\'e}r}, \citenamefont {Maggio-Aprile}, \citenamefont {Jenkins},
  \citenamefont {Risti{\'{c}}}, \citenamefont {Erb}, \citenamefont {Berthod},
  \citenamefont {Fischer},\ and\ \citenamefont {Renner}}]{Bruer2016}%
  \BibitemOpen
  \bibfield  {author} {\bibinfo {author} {\bibfnamefont {J.}~\bibnamefont
  {Bru{\'e}r}}, \bibinfo {author} {\bibfnamefont {I.}~\bibnamefont
  {Maggio-Aprile}}, \bibinfo {author} {\bibfnamefont {N.}~\bibnamefont
  {Jenkins}}, \bibinfo {author} {\bibfnamefont {Z.}~\bibnamefont
  {Risti{\'{c}}}}, \bibinfo {author} {\bibfnamefont {A.}~\bibnamefont {Erb}},
  \bibinfo {author} {\bibfnamefont {C.}~\bibnamefont {Berthod}}, \bibinfo
  {author} {\bibfnamefont {{\O}.}~\bibnamefont {Fischer}}, \ and\ \bibinfo
  {author} {\bibfnamefont {C.}~\bibnamefont {Renner}},\ }\href {\doibase
  10.1038/ncomms11139} {\bibfield  {journal} {\bibinfo  {journal} {Nat. Commun.
  }\ }\textbf {\bibinfo {volume} {7}},\ \bibinfo {pages} {11139}
  (\bibinfo {year} {2016})}\BibitemShut {NoStop}%
\bibitem [{\citenamefont {Ioffe}\ and\ \citenamefont
  {Millis}(2002)}]{PhysRevB.66.094513}%
  \BibitemOpen
  \bibfield  {author} {\bibinfo {author} {\bibfnamefont {L.~B.}\ \bibnamefont
  {Ioffe}}\ and\ \bibinfo {author} {\bibfnamefont {A.~J.}\ \bibnamefont
  {Millis}},\ }\href {\doibase 10.1103/PhysRevB.66.094513} {\bibfield
  {journal} {\bibinfo  {journal} {Phys. Rev. B}\ }\textbf {\bibinfo {volume}
  {66}},\ \bibinfo {pages} {094513} (\bibinfo {year} {2002})}\BibitemShut
  {NoStop}%
\bibitem [{\citenamefont {Wang}\ \emph {et~al.}(2001)\citenamefont {Wang},
  \citenamefont {Han},\ and\ \citenamefont {Lee}}]{PhysRevLett.87.167004}%
  \BibitemOpen
  \bibfield  {author} {\bibinfo {author} {\bibfnamefont {Q.-H.}\ \bibnamefont
  {Wang}}, \bibinfo {author} {\bibfnamefont {J.~H.}\ \bibnamefont {Han}}, \
  and\ \bibinfo {author} {\bibfnamefont {D.-H.}\ \bibnamefont {Lee}},\ }\href
  {\doibase 10.1103/PhysRevLett.87.167004} {\bibfield  {journal} {\bibinfo
  {journal} {Phys. Rev. Lett.}\ }\textbf {\bibinfo {volume} {87}},\ \bibinfo
  {pages} {167004} (\bibinfo {year} {2001})}\BibitemShut {NoStop}%
\bibitem [{\citenamefont {Han}\ and\ \citenamefont
  {Lee}(2000)}]{PhysRevLett.85.1100}%
  \BibitemOpen
  \bibfield  {author} {\bibinfo {author} {\bibfnamefont {J.~H.}\ \bibnamefont
  {Han}}\ and\ \bibinfo {author} {\bibfnamefont {D.-H.}\ \bibnamefont {Lee}},\
  }\href {\doibase 10.1103/PhysRevLett.85.1100} {\bibfield  {journal} {\bibinfo
   {journal} {Phys. Rev. Lett.}\ }\textbf {\bibinfo {volume} {85}},\ \bibinfo
  {pages} {1100} (\bibinfo {year} {2000})}\BibitemShut {NoStop}%
\bibitem [{\citenamefont {Lee}\ \emph {et~al.}(2006)\citenamefont {Lee},
  \citenamefont {Nagaosa},\ and\ \citenamefont {Wen}}]{RevModPhys.78.17}%
  \BibitemOpen
  \bibfield  {author} {\bibinfo {author} {\bibfnamefont {P.~A.}\ \bibnamefont
  {Lee}}, \bibinfo {author} {\bibfnamefont {N.}~\bibnamefont {Nagaosa}}, \ and\
  \bibinfo {author} {\bibfnamefont {X.-G.}\ \bibnamefont {Wen}},\ }\href
  {\doibase 10.1103/RevModPhys.78.17} {\bibfield  {journal} {\bibinfo
  {journal} {Rev. Mod. Phys.}\ }\textbf {\bibinfo {volume} {78}},\ \bibinfo
  {pages} {17} (\bibinfo {year} {2006})}\BibitemShut {NoStop}%
\bibitem [{\citenamefont {Scalapino}(1995)}]{SCALAPINO1995329}%
  \BibitemOpen
  \bibfield  {author} {\bibinfo {author} {\bibfnamefont {D.}~\bibnamefont
  {Scalapino}},\ }\href {\doibase https://doi.org/10.1016/0370-1573(94)00086-I}
  {\bibfield  {journal} {\bibinfo  {journal} {Physics Reports}\ }\textbf
  {\bibinfo {volume} {250}},\ \bibinfo {pages} {329} (\bibinfo {year}
  {1995})}\BibitemShut {NoStop}%
\bibitem [{\citenamefont {Anderson}\ \emph {et~al.}(2004)\citenamefont
  {Anderson}, \citenamefont {Lee}, \citenamefont {Randeria}, \citenamefont
  {Rice}, \citenamefont {Trivedi},\ and\ \citenamefont
  {Zhang}}]{Anderson_2004}%
  \BibitemOpen
  \bibfield  {author} {\bibinfo {author} {\bibfnamefont {P.~W.}\ \bibnamefont
  {Anderson}}, \bibinfo {author} {\bibfnamefont {P.~A.}\ \bibnamefont {Lee}},
  \bibinfo {author} {\bibfnamefont {M.}~\bibnamefont {Randeria}}, \bibinfo
  {author} {\bibfnamefont {T.~M.}\ \bibnamefont {Rice}}, \bibinfo {author}
  {\bibfnamefont {N.}~\bibnamefont {Trivedi}}, \ and\ \bibinfo {author}
  {\bibfnamefont {F.~C.}\ \bibnamefont {Zhang}},\ }\href {\doibase
  10.1088/0953-8984/16/24/r02} {\bibfield  {journal} {\bibinfo  {journal}
  {J. Phys. Condens. Matter }\ }\textbf {\bibinfo {volume} {16}},\
  \bibinfo {pages} {R755} (\bibinfo {year} {2004})}\BibitemShut {NoStop}%
\bibitem [{\citenamefont {Ko}\ \emph {et~al.}(2007)\citenamefont {Ko},
  \citenamefont {Nave},\ and\ \citenamefont {Lee}}]{PhysRevB.76.245113}%
  \BibitemOpen
  \bibfield  {author} {\bibinfo {author} {\bibfnamefont {W.-H.}\ \bibnamefont
  {Ko}}, \bibinfo {author} {\bibfnamefont {C.~P.}\ \bibnamefont {Nave}}, \ and\
  \bibinfo {author} {\bibfnamefont {P.~A.}\ \bibnamefont {Lee}},\ }\href
  {\doibase 10.1103/PhysRevB.76.245113} {\bibfield  {journal} {\bibinfo
  {journal} {Phys. Rev. B}\ }\textbf {\bibinfo {volume} {76}},\ \bibinfo
  {pages} {245113} (\bibinfo {year} {2007})}\BibitemShut {NoStop}%
\bibitem [{\citenamefont {Zhang}\ \emph {et~al.}(1988)\citenamefont {Zhang},
  \citenamefont {Gros}, \citenamefont {Rice},\ and\ \citenamefont
  {Shiba}}]{FCZhang_1988}%
  \BibitemOpen
  \bibfield  {author} {\bibinfo {author} {\bibfnamefont {F.~C.}\ \bibnamefont
  {Zhang}}, \bibinfo {author} {\bibfnamefont {C.}~\bibnamefont {Gros}},
  \bibinfo {author} {\bibfnamefont {T.~M.}\ \bibnamefont {Rice}}, \ and\
  \bibinfo {author} {\bibfnamefont {H.}~\bibnamefont {Shiba}},\ }\href
  {\doibase 10.1088/0953-2048/1/1/009} {\bibfield  {journal} {\bibinfo
  {journal} {Supercond. Sci. Technol.}\ }\textbf {\bibinfo
  {volume} {1}},\ \bibinfo {pages} {36} (\bibinfo {year} {1988})}\BibitemShut
  {NoStop}%
\bibitem [{fnSM()}]{fnSM}%
  \BibitemOpen
  \href@noop {} {}\bibinfo {note} {{See Supplementary Material for details on GRFs, GRF augmented inhomogeneous mean-field theory calculations, order parameter and local density of states from the standard inhomogeneous mean-field theory, and role of the GRFs on emerging Mottness at vortex cores, also including Refs.~\cite{njp.11.055053,prb.84.184511, prl.75.1384,prl.77.566}}}\BibitemShut {NoStop}%
  \bibitem [{\citenamefont {Yang}\ \emph {et~al.}(2009)\citenamefont
  {Yang}, \citenamefont {Chen}, \citenamefont {Rice},\citenamefont {Sigrist}\ and\ \citenamefont
  {Zhang}}]{njp.11.055053}%
  \BibitemOpen
  \bibfield  {author} {\bibinfo {author} {\bibfnamefont {K.-Y.}~\bibnamefont
  {Yang}}, \bibinfo {author} {\bibfnamefont {W.~Q.}~\bibnamefont {Chen}}, \bibinfo {author} {\bibfnamefont {T.~M.}~\bibnamefont {Rice}}, \bibinfo {author} {\bibfnamefont {M.}~\bibnamefont {Sigrist}},\ and\ \bibinfo {author} {\bibfnamefont{F.-C.}~\bibnamefont {Zhang}},\ }\href
  {\doibase 10.1088/1367-2630/11/5/055053} {\bibfield  {journal} {\bibinfo
  {journal} {New. J. Phys.}\ }\textbf {\bibinfo {volume} {11}},\ \bibinfo
  {pages} {055053} (\bibinfo {year} {2009})}\BibitemShut {NoStop}%
\bibitem [{\citenamefont {Christensen}\ \emph {et~al.}(2007)\citenamefont
  {Christensen}, \citenamefont {Hirschfeld}, \citenamefont {Andersen}\ and\ \citenamefont
  {Trivedi}}]{prb.84.184511}%
  \BibitemOpen
  \bibfield  {author} {\bibinfo {author} {\bibfnamefont {R.~B.}~\bibnamefont
  {Christensen}}, \bibinfo {author} {\bibfnamefont {P. ~J.}~\bibnamefont {Hirschfeld}},
  \ and\ \bibinfo {author} {\bibfnamefont {B.~M.}~\bibnamefont {Andersen}},\ }\href
  {\doibase 10.1103/PhysRevB.84.184511} {\bibfield  {journal} {\bibinfo
  {journal} {Phys. Rev. B.}\ }\textbf {\bibinfo {volume} {84}},\ \bibinfo
  {pages} {184511} (\bibinfo {year} {2011})}\BibitemShut {NoStop}%
\bibitem [{\citenamefont {Khomskii}\ \emph {et~al.}(1995)\citenamefont
  {Khomskii}\ and\ \citenamefont
  {Freimuth}}]{prl.75.1384}%
  \BibitemOpen
  \bibfield  {author} {\bibinfo {author} {\bibfnamefont {D.~I.}~\bibnamefont
  {Khomskii}}\ and\ \bibinfo {author} {\bibfnamefont {A.}~\bibnamefont {Freimuth}},\ }\href
  {\doibase 10.1103/PhysRevLett.75.1384} {\bibfield  {journal} {\bibinfo
  {journal} {Phys. Rev. Lett.}\ }\textbf {\bibinfo {volume} {75}},\ \bibinfo
  {pages} {1384} (\bibinfo {year} {1995})}\BibitemShut {NoStop}%
\bibitem [{\citenamefont {Blatter}\ \emph {et~al.}(1996)\citenamefont
  {Blatter}, \citenamefont {Feigel'man}, \citenamefont {Geshkenbein}, \citenamefont {Larkin}\ and\ \citenamefont
  {van Otterlo}}]{prl.77.566}%
  \BibitemOpen
  \bibfield  {author} {\bibinfo {author} {\bibfnamefont {G.}~\bibnamefont
  {Blatter}}, \bibinfo {author} {\bibfnamefont {M.}~\bibnamefont {Feigel'man}}, \bibinfo {author} {\bibfnamefont {V.}~\bibnamefont {Geshkenbein}}, \bibinfo {author} {\bibfnamefont {A.}~\bibnamefont {Larkin}}\ and\ \bibinfo {author} {\bibfnamefont {A.}~\bibnamefont {van Otterlo}},\ }\href
  {\doibase 10.1103/PhysRevLett.77.566} {\bibfield  {journal} {\bibinfo
  {journal} {Phys. Rev. Lett.}\ }\textbf {\bibinfo {volume} {77}},\ \bibinfo
  {pages} {566} (\bibinfo {year} {1996})}\BibitemShut {NoStop}%
\bibitem [{\citenamefont {Sensarma}\ \emph {et~al.}(2007)\citenamefont
  {Sensarma}, \citenamefont {Randeria},\ and\ \citenamefont
  {Trivedi}}]{PhysRevLett.98.027004}%
  \BibitemOpen
  \bibfield  {author} {\bibinfo {author} {\bibfnamefont {R.}~\bibnamefont
  {Sensarma}}, \bibinfo {author} {\bibfnamefont {M.}~\bibnamefont {Randeria}},
  \ and\ \bibinfo {author} {\bibfnamefont {N.}~\bibnamefont {Trivedi}},\ }\href
  {\doibase 10.1103/PhysRevLett.98.027004} {\bibfield  {journal} {\bibinfo
  {journal} {Phys. Rev. Lett.}\ }\textbf {\bibinfo {volume} {98}},\ \bibinfo
  {pages} {027004} (\bibinfo {year} {2007})}\BibitemShut {NoStop}%
  \bibitem [{\citenamefont {Himeda}\ and\ \citenamefont
  {Ogata}(1999)}]{PhysRevB.60.R9935}%
  \BibitemOpen
  \bibfield  {author} {\bibinfo {author} {\bibfnamefont {A.}~\bibnamefont
  {Himeda}}\ and\ \bibinfo {author} {\bibfnamefont {M.}~\bibnamefont {Ogata}},\
  }\href {\doibase 10.1103/PhysRevB.60.R9935} {\bibfield  {journal} {\bibinfo
  {journal} {Phys. Rev. B}\ }\textbf {\bibinfo {volume} {60}},\ \bibinfo
  {pages} {R9935} (\bibinfo {year} {1999})}\BibitemShut {NoStop}%
\bibitem [{\citenamefont {Paramekanti}\ \emph {et~al.}(2001)\citenamefont
  {Paramekanti}, \citenamefont {Randeria},\ and\ \citenamefont
  {Trivedi}}]{PhysRevLett.87.217002}%
  \BibitemOpen
  \bibfield  {author} {\bibinfo {author} {\bibfnamefont {A.}~\bibnamefont
  {Paramekanti}}, \bibinfo {author} {\bibfnamefont {M.}~\bibnamefont
  {Randeria}}, \ and\ \bibinfo {author} {\bibfnamefont {N.}~\bibnamefont
  {Trivedi}},\ }\href {\doibase 10.1103/PhysRevLett.87.217002} {\bibfield
  {journal} {\bibinfo  {journal} {Phys. Rev. Lett.}\ }\textbf {\bibinfo
  {volume} {87}},\ \bibinfo {pages} {217002} (\bibinfo {year}
  {2001})}\BibitemShut {NoStop}%
\bibitem [{\citenamefont {Norman}\ \emph {et~al.}(1995)\citenamefont {Norman},
  \citenamefont {Randeria}, \citenamefont {Ding},\ and\ \citenamefont
  {Campuzano}}]{PhysRevB.52.615}%
  \BibitemOpen
  \bibfield  {author} {\bibinfo {author} {\bibfnamefont {M.~R.}\ \bibnamefont
  {Norman}}, \bibinfo {author} {\bibfnamefont {M.}~\bibnamefont {Randeria}},
  \bibinfo {author} {\bibfnamefont {H.}~\bibnamefont {Ding}}, \ and\ \bibinfo
  {author} {\bibfnamefont {J.~C.}\ \bibnamefont {Campuzano}},\ }\href {\doibase
  10.1103/PhysRevB.52.615} {\bibfield  {journal} {\bibinfo  {journal} {Phys.
  Rev. B}\ }\textbf {\bibinfo {volume} {52}},\ \bibinfo {pages} {615} (\bibinfo
  {year} {1995})}\BibitemShut {NoStop}%
\bibitem [{\citenamefont {Garg}\ \emph {et~al.}(2008)\citenamefont {Garg},
  \citenamefont {Randeria},\ and\ \citenamefont {Trivedi}}]{Garg2008}%
  \BibitemOpen
  \bibfield  {author} {\bibinfo {author} {\bibfnamefont {A.}~\bibnamefont
  {Garg}}, \bibinfo {author} {\bibfnamefont {M.}~\bibnamefont {Randeria}}, \
  and\ \bibinfo {author} {\bibfnamefont {N.}~\bibnamefont {Trivedi}},\ }\href
  {\doibase 10.1038/nphys1026} {\bibfield  {journal} {\bibinfo  {journal}
  {Nat. Phys.}\ }\textbf {\bibinfo {volume} {4}},\ \bibinfo {pages} {762}
  (\bibinfo {year} {2008})}\BibitemShut {NoStop}%
\bibitem [{\citenamefont {Fukushima}(2008)}]{PhysRevB.78.115105}%
  \BibitemOpen
  \bibfield  {author} {\bibinfo {author} {\bibfnamefont {N.}~\bibnamefont
  {Fukushima}},\ }\href {\doibase 10.1103/PhysRevB.78.115105} {\bibfield
  {journal} {\bibinfo  {journal} {Phys. Rev. B}\ }\textbf {\bibinfo {volume}
  {78}},\ \bibinfo {pages} {115105} (\bibinfo {year} {2008})}\BibitemShut
  {NoStop}%
\bibitem [{\citenamefont {Chakraborty}\ \emph {et~al.}(2017)\citenamefont
  {Chakraborty}, \citenamefont {Sensarma},\ and\ \citenamefont
  {Ghosal}}]{PhysRevB.95.014516}%
  \BibitemOpen
  \bibfield  {author} {\bibinfo {author} {\bibfnamefont {D.}~\bibnamefont
  {Chakraborty}}, \bibinfo {author} {\bibfnamefont {R.}~\bibnamefont
  {Sensarma}}, \ and\ \bibinfo {author} {\bibfnamefont {A.}~\bibnamefont
  {Ghosal}},\ }\href {\doibase 10.1103/PhysRevB.95.014516} {\bibfield
  {journal} {\bibinfo  {journal} {Phys. Rev. B}\ }\textbf {\bibinfo {volume}
  {95}},\ \bibinfo {pages} {014516} (\bibinfo {year} {2017})}\BibitemShut
  {NoStop}%
\bibitem [{\citenamefont {Caroli}\ \emph {et~al.}(1964)\citenamefont {Caroli},
  \citenamefont {Gennes},\ and\ \citenamefont {Matricon}}]{CDGM_PhysLett9_307}%
  \BibitemOpen
  \bibfield  {author} {\bibinfo {author} {\bibfnamefont {C.}~\bibnamefont
  {Caroli}}, \bibinfo {author} {\bibfnamefont {P.~D.}\ \bibnamefont {Gennes}},
  \ and\ \bibinfo {author} {\bibfnamefont {J.}~\bibnamefont {Matricon}},\
  }\href {\doibase https://doi.org/10.1016/0031-9163(64)90375-0} {\bibfield
  {journal} {\bibinfo  {journal} {Phys. Lett.}\ }\textbf {\bibinfo {volume}
  {9}},\ \bibinfo {pages} {307} (\bibinfo {year} {1964})}\BibitemShut {NoStop}%
\bibitem [{\citenamefont {Hess}\ \emph {et~al.}(1989)\citenamefont {Hess},
  \citenamefont {Robinson}, \citenamefont {Dynes}, \citenamefont {Valles},\
  and\ \citenamefont {Waszczak}}]{PhysRevLett.62.214}%
  \BibitemOpen
  \bibfield  {author} {\bibinfo {author} {\bibfnamefont {M.}~\bibnamefont
  {Chen}}, \bibinfo {author} {\bibfnamefont {X.}~\bibnamefont {Chen}}, \bibinfo
  {author} {\bibfnamefont {H.}~\bibnamefont {Yang}}, \bibinfo {author}
  {\bibfnamefont {Z.}~\bibnamefont {Du}}, \bibinfo {author} {\bibfnamefont
  {X.}~\bibnamefont {Zhu}}, \bibinfo {author} {\bibfnamefont {E.}~\bibnamefont
  {Wang}}, \ and\ \bibinfo {author} {\bibfnamefont {H.-H.}\ \bibnamefont
  {Wen}},\ }\href {\doibase 10.1038/s41467-018-03404-8} {\bibfield  {journal}
  {\bibinfo  {journal} {Nat. Commun.}\ }\textbf {\bibinfo {volume}
  {9}},\ \bibinfo {pages} {970} (\bibinfo {year} {2018})}\BibitemShut {NoStop}%
\bibitem [{\citenamefont {Datta}\ \emph {et~al.}(2019)\citenamefont {Datta},
  \citenamefont {Yang},\ and\ \citenamefont {Ghosal}}]{PhysRevB.100.035114}%
  \BibitemOpen
  \bibfield  {author} {\bibinfo {author} {\bibfnamefont {A.}~\bibnamefont
  {Datta}}, \bibinfo {author} {\bibfnamefont {K.}~\bibnamefont {Yang}}, \ and\
  \bibinfo {author} {\bibfnamefont {A.}~\bibnamefont {Ghosal}},\ }\href
  {\doibase 10.1103/PhysRevB.100.035114} {\bibfield  {journal} {\bibinfo
  {journal} {Phys. Rev. B}\ }\textbf {\bibinfo {volume} {100}},\ \bibinfo
  {pages} {035114} (\bibinfo {year} {2019})}\BibitemShut {NoStop}%
\bibitem [{\citenamefont {Anderson}(2006)}]{PhysRevLett.96.017001}%
  \BibitemOpen
  \bibfield  {author} {\bibinfo {author} {\bibfnamefont {P.~W.}\ \bibnamefont
  {Anderson}},\ }\href {\doibase 10.1103/PhysRevLett.96.017001} {\bibfield
  {journal} {\bibinfo  {journal} {Phys. Rev. Lett.}\ }\textbf {\bibinfo
  {volume} {96}},\ \bibinfo {pages} {017001} (\bibinfo {year}
  {2006})}\BibitemShut {NoStop}%
\bibitem [{\citenamefont {Scalapino}\ \emph {et~al.}(1993)\citenamefont
  {Scalapino}, \citenamefont {White},\ and\ \citenamefont
  {Zhang}}]{PhysRevB.47.7995}%
  \BibitemOpen
  \bibfield  {author} {\bibinfo {author} {\bibfnamefont {D.~J.}\ \bibnamefont
  {Scalapino}}, \bibinfo {author} {\bibfnamefont {S.~R.}\ \bibnamefont
  {White}}, \ and\ \bibinfo {author} {\bibfnamefont {S.}~\bibnamefont
  {Zhang}},\ }\href {\doibase 10.1103/PhysRevB.47.7995} {\bibfield  {journal}
  {\bibinfo  {journal} {Phys. Rev. B}\ }\textbf {\bibinfo {volume} {47}},\
  \bibinfo {pages} {7995} (\bibinfo {year} {1993})}\BibitemShut {NoStop}%
\bibitem [{\citenamefont {Wen}\ \emph {et~al.}(2003)\citenamefont {Wen},
  \citenamefont {Yang}, \citenamefont {Li}, \citenamefont {Zeng}, \citenamefont
  {Soukiassian}, \citenamefont {Si},\ and\ \citenamefont {Xi}}]{Wen_2003}%
  \BibitemOpen
  \bibfield  {author} {\bibinfo {author} {\bibfnamefont {H.~H.}\ \bibnamefont
  {Wen}}, \bibinfo {author} {\bibfnamefont {H.~P.}\ \bibnamefont {Yang}},
  \bibinfo {author} {\bibfnamefont {S.~L.}\ \bibnamefont {Li}}, \bibinfo
  {author} {\bibfnamefont {X.~H.}\ \bibnamefont {Zeng}}, \bibinfo {author}
  {\bibfnamefont {A.~A.}\ \bibnamefont {Soukiassian}}, \bibinfo {author}
  {\bibfnamefont {W.~D.}\ \bibnamefont {Si}}, \ and\ \bibinfo {author}
  {\bibfnamefont {X.~X.}\ \bibnamefont {Xi}},\ }\href {\doibase
  10.1209/epl/i2003-00627-1} {\bibfield  {journal} {\bibinfo  {journal}
  {Europhysics Letters ({EPL})}\ }\textbf {\bibinfo {volume} {64}},\ \bibinfo
  {pages} {790} (\bibinfo {year} {2003})}\BibitemShut {NoStop}%
\bibitem [{\citenamefont {Ramshaw}\ \emph {et~al.}(2012)\citenamefont
  {Ramshaw}, \citenamefont {Day}, \citenamefont {Vignolle}, \citenamefont
  {LeBoeuf}, \citenamefont {Dosanjh}, \citenamefont {Proust}, \citenamefont
  {Taillefer}, \citenamefont {Liang}, \citenamefont {Hardy},\ and\
  \citenamefont {Bonn}}]{PhysRevB.86.174501}%
  \BibitemOpen
  \bibfield  {author} {\bibinfo {author} {\bibfnamefont {B.~J.}\ \bibnamefont
  {Ramshaw}}, \bibinfo {author} {\bibfnamefont {J.}~\bibnamefont {Day}},
  \bibinfo {author} {\bibfnamefont {B.}~\bibnamefont {Vignolle}}, \bibinfo
  {author} {\bibfnamefont {D.}~\bibnamefont {LeBoeuf}}, \bibinfo {author}
  {\bibfnamefont {P.}~\bibnamefont {Dosanjh}}, \bibinfo {author} {\bibfnamefont
  {C.}~\bibnamefont {Proust}}, \bibinfo {author} {\bibfnamefont
  {L.}~\bibnamefont {Taillefer}}, \bibinfo {author} {\bibfnamefont
  {R.}~\bibnamefont {Liang}}, \bibinfo {author} {\bibfnamefont {W.~N.}\
  \bibnamefont {Hardy}}, \ and\ \bibinfo {author} {\bibfnamefont {D.~A.}\
  \bibnamefont {Bonn}},\ }\href {\doibase 10.1103/PhysRevB.86.174501}
  {\bibfield  {journal} {\bibinfo  {journal} {Phys. Rev. B}\ }\textbf {\bibinfo
  {volume} {86}},\ \bibinfo {pages} {174501} (\bibinfo {year}
  {2012})}\BibitemShut {NoStop}%
  \bibitem [{\citenamefont {Wang}\ and\ \citenamefont {Wen}(2008)}]{Wang2008}%
  \BibitemOpen
  \bibfield  {author} {\bibinfo {author} {\bibfnamefont {Y.}~\bibnamefont
  {Wang}}\ and\ \bibinfo {author} {\bibfnamefont {H.-H.}\ \bibnamefont {Wen}},\
  }\href {\doibase 10.1209/0295-5075/81/57007} {\bibfield  {journal} {\bibinfo
  {journal} {{Europhysics Letters} (EPL)}\ }\textbf {\bibinfo {volume} {81}},\
  \bibinfo {pages} {57007} (\bibinfo {year} 
  {2008})}\BibitemShut {NoStop}%
\bibitem [{\citenamefont {Sahu}\ \emph {et~al.}(2022)\citenamefont {Sahu},
  \citenamefont {Mandal}, \citenamefont {Ghosh}, \citenamefont {Deshmukh},\
  and\ \citenamefont {Singh}}]{doi:10.1021/acs.nanolett.1c04688}%
  \BibitemOpen
  \bibfield  {author} {\bibinfo {author} {\bibfnamefont {S.~K.}\ \bibnamefont
  {Sahu}}, \bibinfo {author} {\bibfnamefont {S.}~\bibnamefont {Mandal}},
  \bibinfo {author} {\bibfnamefont {S.}~\bibnamefont {Ghosh}}, \bibinfo
  {author} {\bibfnamefont {M.~M.}\ \bibnamefont {Deshmukh}}, \ and\ \bibinfo
  {author} {\bibfnamefont {V.}~\bibnamefont {Singh}},\ }\href {\doibase
  10.1021/acs.nanolett.1c04688} {\bibfield  {journal} {\bibinfo  {journal}
  {Nano Letters}\ }\textbf {\bibinfo {volume} {22}},\ \bibinfo {pages} {1665}
  (\bibinfo {year} {2022})}\BibitemShut {NoStop}%
\bibitem [{\citenamefont {Grissonnanche}\ \emph {et~al.}(2022)\citenamefont {Cyr-Choini{\`e}re},
  \citenamefont {Lalibert{\'e}}, \citenamefont {Ren{\'e} de Cotret}, \citenamefont {Deshmukh},\
  and\ \citenamefont {Juneau-Fecteau}}]{Grissonnanche2014}%
  \BibitemOpen
  \bibfield  {author} {\bibinfo {author} {\bibfnamefont {G.}\ \bibnamefont
  {Grissonnanche}}, \bibinfo {author} {\bibfnamefont {O.}~\bibnamefont {Cyr-Choini{\`e}re}},
  \bibinfo {author} {\bibfnamefont {F.}~\bibnamefont {Lalibert{\'e}}}, \bibinfo
  {author} {\bibfnamefont {S.}\ \bibnamefont {Ren{\'e} de Cotret}}, \bibinfo
  {author} {\bibfnamefont {S.}\ \bibnamefont {Juneau-Fecteau}}, \bibinfo
  {author} {\bibfnamefont {S.}\ \bibnamefont {Dufour-Beaus{\'e}jour}}, \bibinfo
  {author} {\bibfnamefont {M.-{\`E}}\ \bibnamefont {Delage}}, \bibinfo
  {author} {\bibfnamefont {D.}\ \bibnamefont {LeBoeuf}}, \bibinfo
  {author} {\bibfnamefont {J.}\ \bibnamefont {Chang}}, \ \bibinfo
  {author} {\bibfnamefont {B. J.}\ \bibnamefont {Ramshaw}}, \ \bibinfo
  {author} {\bibfnamefont {D. A.}\ \bibnamefont {Bonn}}, \bibinfo
  {author} {\bibfnamefont {W. N.}\ \bibnamefont {Hardy}}, \bibinfo
  {author} {\bibfnamefont {R.}\ \bibnamefont {Liang}}, \bibinfo
  {author} {\bibfnamefont {S.}\ \bibnamefont {Adachi}}, \bibinfo
  {author} {\bibfnamefont {N. E.}\ \bibnamefont {Hussey}}, \bibinfo
  {author} {\bibfnamefont {B.}\ \bibnamefont {Vignolle}}, \bibinfo
  {author} {\bibfnamefont {C.}\ \bibnamefont {Proust}}, \bibinfo
  {author} {\bibfnamefont {M.}\ \bibnamefont {Sutherland}}, \bibinfo
  {author} {\bibfnamefont {J.-H.}\ \bibnamefont {Park}}, \bibinfo
  {author} {\bibfnamefont {D.}\ \bibnamefont {Graf}}, \bibinfo
  {author} {\bibfnamefont {N.}\ \bibnamefont {Doiron-Leyraud}}\ and\ \bibinfo
  {author} {\bibfnamefont {L.}\ \bibnamefont {Taillefer}},\ }\href {https://doi.org/10.1038/ncomms4280} {\bibfield  {journal} {\bibinfo  {journal}
  {Nat. Commun.}\ }\textbf {\bibinfo {volume} {5}},\ \bibinfo {pages} {3280}
  (\bibinfo {year} {2014})}\BibitemShut {NoStop}%
\bibitem [{\citenamefont {Sprau}\ \emph {et~al.}(2017)\citenamefont {Sprau},
  \citenamefont {Kostin}, \citenamefont {Kreisel}, \citenamefont {Böhmer},
  \citenamefont {Taufour}, \citenamefont {Canfield}, \citenamefont {Mukherjee},
  \citenamefont {Hirschfeld}, \citenamefont {Andersen},\ and\ \citenamefont
  {Davis}}]{doi:10.1126/science.aal1575}%
  \BibitemOpen
  \bibfield  {author} {\bibinfo {author} {\bibfnamefont {P.~O.}\ \bibnamefont
  {Sprau}}, \bibinfo {author} {\bibfnamefont {A.}~\bibnamefont {Kostin}},
  \bibinfo {author} {\bibfnamefont {A.}~\bibnamefont {Kreisel}}, \bibinfo
  {author} {\bibfnamefont {A.~E.}\ \bibnamefont {Böhmer}}, \bibinfo {author}
  {\bibfnamefont {V.}~\bibnamefont {Taufour}}, \bibinfo {author} {\bibfnamefont
  {P.~C.}\ \bibnamefont {Canfield}}, \bibinfo {author} {\bibfnamefont
  {S.}~\bibnamefont {Mukherjee}}, \bibinfo {author} {\bibfnamefont {P.~J.}\
  \bibnamefont {Hirschfeld}}, \bibinfo {author} {\bibfnamefont {B.~M.}\
  \bibnamefont {Andersen}}, \ and\ \bibinfo {author} {\bibfnamefont {J.~C.~S.}\
  \bibnamefont {Davis}},\ }\href {\doibase 10.1126/science.aal1575} {\bibfield
  {journal} {\bibinfo  {journal} {Science}\ }\textbf {\bibinfo {volume}
  {357}},\ \bibinfo {pages} {75} (\bibinfo {year} {2017})}\BibitemShut
  {NoStop}%
\bibitem [{\citenamefont {de' Medici}\ \emph {et~al.}(2014)\citenamefont {de'
  Medici}, \citenamefont {Giovannetti},\ and\ \citenamefont
  {Capone}}]{PhysRevLett.112.177001}%
  \BibitemOpen
  \bibfield  {author} {\bibinfo {author} {\bibfnamefont {L.}~\bibnamefont {de'
  Medici}}, \bibinfo {author} {\bibfnamefont {G.}~\bibnamefont {Giovannetti}},
  \ and\ \bibinfo {author} {\bibfnamefont {M.}~\bibnamefont {Capone}},\ }\href
  {\doibase 10.1103/PhysRevLett.112.177001} {\bibfield  {journal} {\bibinfo
  {journal} {Phys. Rev. Lett.}\ }\textbf {\bibinfo {volume} {112}},\ \bibinfo
  {pages} {177001} (\bibinfo {year} {2014})}\BibitemShut {NoStop}%
\bibitem [{\citenamefont {Wang}\ \emph {et~al.}(2018)\citenamefont {Wang},
  \citenamefont {Kong}, \citenamefont {Fan}, \citenamefont {Chen},
  \citenamefont {Zhu}, \citenamefont {Liu}, \citenamefont {Cao}, \citenamefont
  {Sun}, \citenamefont {Du}, \citenamefont {Schneeloch}, \citenamefont {Zhong},
  \citenamefont {Gu}, \citenamefont {Fu}, \citenamefont {Ding},\ and\
  \citenamefont {Gao}}]{doi:10.1126/science.aao1797}%
  \BibitemOpen
  \bibfield  {author} {\bibinfo {author} {\bibfnamefont {D.}~\bibnamefont
  {Wang}}, \bibinfo {author} {\bibfnamefont {L.}~\bibnamefont {Kong}}, \bibinfo
  {author} {\bibfnamefont {P.}~\bibnamefont {Fan}}, \bibinfo {author}
  {\bibfnamefont {H.}~\bibnamefont {Chen}}, \bibinfo {author} {\bibfnamefont
  {S.}~\bibnamefont {Zhu}}, \bibinfo {author} {\bibfnamefont {W.}~\bibnamefont
  {Liu}}, \bibinfo {author} {\bibfnamefont {L.}~\bibnamefont {Cao}}, \bibinfo
  {author} {\bibfnamefont {Y.}~\bibnamefont {Sun}}, \bibinfo {author}
  {\bibfnamefont {S.}~\bibnamefont {Du}}, \bibinfo {author} {\bibfnamefont
  {J.}~\bibnamefont {Schneeloch}}, \bibinfo {author} {\bibfnamefont
  {R.}~\bibnamefont {Zhong}}, \bibinfo {author} {\bibfnamefont
  {G.}~\bibnamefont {Gu}}, \bibinfo {author} {\bibfnamefont {L.}~\bibnamefont
  {Fu}}, \bibinfo {author} {\bibfnamefont {H.}~\bibnamefont {Ding}}, \ and\
  \bibinfo {author} {\bibfnamefont {H.-J.}\ \bibnamefont {Gao}},\ }\href
  {\doibase 10.1126/science.aao1797} {\bibfield  {journal} {\bibinfo  {journal}
  {Science}\ }\textbf {\bibinfo {volume} {362}},\ \bibinfo {pages} {333}
  (\bibinfo {year} {2018})}\BibitemShut {NoStop}%
\bibitem [{\citenamefont {Cao}\ \emph {et~al.}(2018)\citenamefont {Cao},
  \citenamefont {Fatemi}, \citenamefont {Demir}, \citenamefont {Fang},
  \citenamefont {Tomarken}, \citenamefont {Luo}, \citenamefont
  {Sanchez-Yamagishi}, \citenamefont {Watanabe}, \citenamefont {Taniguchi},
  \citenamefont {Kaxiras}, \citenamefont {Ashoori},\ and\ \citenamefont
  {Jarillo-Herrero}}]{Cao2018}%
  \BibitemOpen
  \bibfield  {author} {\bibinfo {author} {\bibfnamefont {Y.}~\bibnamefont
  {Cao}}, \bibinfo {author} {\bibfnamefont {V.}~\bibnamefont {Fatemi}},
  \bibinfo {author} {\bibfnamefont {A.}~\bibnamefont {Demir}}, \bibinfo
  {author} {\bibfnamefont {S.}~\bibnamefont {Fang}}, \bibinfo {author}
  {\bibfnamefont {S.~L.}\ \bibnamefont {Tomarken}}, \bibinfo {author}
  {\bibfnamefont {J.~Y.}\ \bibnamefont {Luo}}, \bibinfo {author} {\bibfnamefont
  {J.~D.}\ \bibnamefont {Sanchez-Yamagishi}}, \bibinfo {author} {\bibfnamefont
  {K.}~\bibnamefont {Watanabe}}, \bibinfo {author} {\bibfnamefont
  {T.}~\bibnamefont {Taniguchi}}, \bibinfo {author} {\bibfnamefont
  {E.}~\bibnamefont {Kaxiras}}, \bibinfo {author} {\bibfnamefont {R.~C.}\
  \bibnamefont {Ashoori}}, \ and\ \bibinfo {author} {\bibfnamefont
  {P.}~\bibnamefont {Jarillo-Herrero}},\ }\href {\doibase 10.1038/nature26154}
  {\bibfield  {journal} {\bibinfo  {journal} {Nature}\ }\textbf {\bibinfo
  {volume} {556}},\ \bibinfo {pages} {80} (\bibinfo {year} {2018})}\BibitemShut
  {NoStop}%
\bibitem [{\citenamefont {Ganguli}\ \emph {et~al.}(2016)\citenamefont
  {Ganguli}, \citenamefont {Singh}, \citenamefont {Ganguly}, \citenamefont
  {Bagwe}, \citenamefont {Thamizhavel},\ and\ \citenamefont
  {Raychaudhuri}}]{Ganguli_2016}%
  \BibitemOpen
  \bibfield  {author} {\bibinfo {author} {\bibfnamefont {S.~C.}\ \bibnamefont
  {Ganguli}}, \bibinfo {author} {\bibfnamefont {H.}~\bibnamefont {Singh}},
  \bibinfo {author} {\bibfnamefont {R.}~\bibnamefont {Ganguly}}, \bibinfo
  {author} {\bibfnamefont {V.}~\bibnamefont {Bagwe}}, \bibinfo {author}
  {\bibfnamefont {A.}~\bibnamefont {Thamizhavel}}, \ and\ \bibinfo {author}
  {\bibfnamefont {P.}~\bibnamefont {Raychaudhuri}},\ }\href {\doibase
  10.1088/0953-8984/28/16/165701} {\bibfield  {journal} {\bibinfo  {journal}
  {J. Phys. Condens. Matter}\ }\textbf {\bibinfo {volume} {28}},\
  \bibinfo {pages} {165701} (\bibinfo {year} {2016})}\BibitemShut {NoStop}%
\end{thebibliography}

\begin{thebibliography}{11}
\bibitem{KoNaveLee} W.-H. Ko, C. P. Nave, and P. A. Lee, Phys. Rev. B \textbf{76}, 245113 (2007).
\bibitem{Zhang_1988} F. C. Zhang, C. Gros, T. M. Rice, and H. Shiba, Supercond. Sci. Technol. \textbf{1}, 36 (1988).
\bibitem{Yang_2009} K.-Y. Yang et. al., New J. Phys. \textbf{11}, 055053 (2009).
\bibitem{Christensen_2011} R. B. Christensen, P. J. Hirschfeld, and B. M. Andersen, Phys. Rev. B \textbf{84}, 184511 (2011).
\bibitem{Garg_2008} A. Garg, M. Randeria, and N. Trivedi, Nat. Phys. \textbf{4}, 762 (2008).
\bibitem{Ghosal_2002} A. Ghosal, C. Kallin, and A. J. Berlinsky, Phys. Rev. B \textbf{66}, 214502 (2002).
\bibitem{Wang_1995} Y. Wang and A. H. MacDonald, Phys. Rev. B \textbf{52}, R3876 (1995).
\bibitem{Khomskii_1995} D. I. Khomskii and A. Freimuth, Phys. Rev. Lett. \textbf{75}, 1384 (1995).
\bibitem{Blatter_1996} G. Blatter, M. Feigel'man, V. Geshkenbein and A. Larkin, and A. van Otterlo, Phys. Rev. Lett. \textbf{77}, 566 (1996).
\end{thebibliography}
\end{document}